\documentclass[floatfix,showpacs,superscriptaddress,prl]{revtex4-2}
\usepackage{amsbsy,amssymb,amsmath,bm}
\usepackage{float}
\usepackage{epsf}
\usepackage{graphicx,color}
\usepackage[caption = false]{subfig}
\usepackage{amsfonts}
\usepackage{amssymb}
\usepackage{amsmath}
\usepackage[T1]{fontenc}
\usepackage{xspace}
\usepackage{epsfig}
\usepackage{ctable}
\usepackage{enumitem}
\newcommand{\rrangle}{>\kern-1.2ex~>\xspace}
\definecolor{red}{rgb}{1,0,0}           % Standard colours red, green, blue
\definecolor{green}{rgb}{0,1,0}
\definecolor{blue}{rgb}{0,0,1}
\definecolor{darkblue}{rgb}{0,0,0.5}
\definecolor{lightblue}{rgb}{.5,.5,1}
\definecolor{lightgray}{gray}{.87}          % How you can define your own greys
\definecolor{Dark}{gray}{.20}
\definecolor{pink}{rgb}{.95,0.82,0.92}  % How you can define your own colours
\definecolor{yellow}{rgb}{1,1,0}
\definecolor{lightyellow}{rgb}{1,1,.5}
\definecolor{purple}{rgb}{0.7,0,0.85}
\definecolor{darkgreen}{rgb}{0,0.5,0}
\definecolor{orange}{rgb}{0.8,0.2,0.2}

\def \be {\begin{equation}}
\def \ee {\end{equation}}
\def \bea {\begin{eqnarray}}
\def \eea {\end{eqnarray}}
\def \bse {\begin{subequations}}
\def \ese {\end{subequations}}
\def \bde {\begin{description}}
\def \ede {\end{description}}
\def \bee {\begin{enumerate}}
\def \eee {\end{enumerate}}
\def \nn {\nonumber}

\def \at {\tilde{a}}
\def \bt {\tilde{b}}
\def \kt {\tilde{k}}
\def \kh {\hat{k}}
\def \Kh {\hat{K}}

\def \a {\alpha}
\def \b {\beta}

\def \g {\gamma}

\def \n {\nu}
\def \k {\kappa}

\def \Om {\Omega}

\def \Omh {\hat{\Omega}}

\begin{document}

\title{SSH coupled-spring systems }

\author{Jie-Ying Kuo\footnote{e-mail address: kuojieying@gmail.com}}
\affiliation{\textit{Department of Physics, National Central University, Taoyuan  320317, Taiwan}}
\author{Tsung-Yen Lee\footnote{e-mail address: lsjh.103.707.25@gmail.com}}
\affiliation{\textit{Department of Electrical Engineering, National Chiayi University, Chiayi 60004, Taiwan}}
\author{Yi-Chia Chiu\footnote{e-mail address: justin910306@gmail.com}}
\affiliation{\textit{Department of Physics, National Tsing Hua University, Hsinchu 30013, Taiwan}}
\author{Sheng-Rong Liao\footnote{e-mail address: b11202037@ntu.edu.tw}}
\affiliation{\textit{Physics Department, National Taiwan University, Taipei 10617, Taiwan}}
\author{Hsien-chung Kao\footnote{e-mail address: hckao@phy.ntnu.edu.tw}}
\affiliation{\textit{Physics Department, National Taiwan Normal University, Taipei 11677, Taiwan}}

\date{\today }

\begin{abstract}
It is known that there is also a topological phase in the SSH coupled-spring system with the fixed-end boundary conditions. When this is the case, there would exist edge modes on its boundaries.  In contrast, if the system satisfies the free-end boundary conditions, there is no edge mode, even if it is the topological phase. We show that by varying the force constant of the spring by the boundary in such a system, edge modes would generally appear independent of whether the bulk of the system is in the topological or trivial phases. Moreover, edge modes could exist even if the system satisfies the free-end boundary conditions.
\end{abstract}

%%\pacs{74.20.-z, 74.78.-w, 74.25.F-,71.10.Pm}
\maketitle

\subsection{I. Introduction}

Ever since topological materials were discovered, they have attracted a lot of attention in the community of condensed matter physicists~\cite{Review1, Review2}.  Bulk-boundary correspondence (BBC) is one of the most remarkable properties of topological insulators and superconductors.  A ``periodic table'' has been put forward that classify them according to the symmetries and dimensionalities of the system~\cite{Periodic table1, Periodic table2, Periodic table3}. From the Bloch bands of a system, one can define topological invariants that may be used to tell whether the system is in the topological or trivial phases.  It, in turn, determines the number of edge states that appear on the boundaries of the system.

The Su–Schrieffer–Heeger (SSH) model is a model for dimerized polyacetylene chains, which is considered to be the simplest topological insulator~\cite{SSH}. Since it is relatively easy to realize, such a system has also been studied by experimentalists. See Ref.~\cite{SSH-QD} and references therein for details. It is known that the relevant topological invariant in this system is the one-dimensional (1D) winding number $\n$, which may be used to predict the number of edge states on the boundaries. The 1D winding number is equivalent to the Zak phase, $\g$~\cite{Zak}, which is analogous to the Berry phase~\cite{Berry} and has been measured directly using a dimerized optical lattice~\cite{Zak phase-OL}.

Because of its simplicity, the SSH model has been used as a prototype to investigate various aspects of topological insulators.  First, one may add an on-site energy term and extend it to the Rice-Mele model~\cite{Rice-Mele}. This model is then used to relate the SSH model to a Chern insulator via a charge-pumping process~\cite{Thouless, polarization1, polarization2}.  Moreover, one may introduce third-nearest-neighbor hopping amplitudes in the system so that there would be topological phases with higher winding numbers. They are usually referred to as the extended SSH models, and they may be used to investigate the ambiguity in defining the Zak phase~\cite{Kudin, Rhim, Ext-SSH}. One may also add in such a system more ``atoms'' in a unit cell and generalize it to multi-band SSH models~\cite{Multi-band1, Multi-band2, Multi-band3}.  As a result, one must use holonomies to describe the topology of these systems~\cite{Holonomy, W-loop}.  This is effectively the non-Abelian generalization of the Berry phase.

Topological phonon modes were first investigated in microtubes and filamentary structures in Ref.~\cite{Prodan2009, Prodan2011}. It was then pointed out in Ref.~\cite{KaneLub2014} that topological property also exists in classical mechanical systems, whose Hamiltonians may be mapped to those of quantum electronic systems. As a result, they may be classified in a way similar to the topological insulators and superconductors~\cite{KaneLub2014, Huber2016}.  Soon after the publication of Ref.~\cite{KaneLub2014}, chains of rigid bars connected by linkages were used to build a mechanical analog of the SSH model~\cite{ChenVit2014}, which we will refer to as the SSH coupled-spring system hereafter. The authors use the system to study the mid-gap states, the counterparts of the edge states in the SSH model, that are localized at the boundaries. Since then, there have been extensive studies along this line~[28–46].

In Ref.~\cite{Grundmann2020}, the author investigates in detail the SSH coupled-spring system satisfying the fixed-end and free-end boundary conditions (BCs). It has been shown that when the system satisfies the free-end BCs, then there would not be any edge states even if the system is in the topological phase. Moreover, they modify the force constant of the spring attached to the boundary sites and study its effect on the edge states. In this work, we take up a similar task. We mainly focus on the edge states, and we try to carry out analytic calculations as much as we can. For this purpose, we use semi-infinite chains to simplify our analysis.  The rest of the paper is organized in the following way. In Sec. II, we first use the SSH coupled-spring system satisfying the fixed-end BCs to explain how we may analyze such systems. It is known that edge states exist on the boundaries of the system when it is in the topological phase. In particular, we show how to find the edge states analytically in a semi-infinite chain.  In Sec. III,  we show analytically that there is no edge state if the system satisfies the free-end BCs. As the existence of edge states depends sensitively on the boundary conditions, we then study in detail whether edge states would exist if we modify the force constant of the spring between the sites by the boundary in a semi-infinite chain in Sec. IV.  Both fixed-end and free-end BCs are considered, and we find that edge modes could exist even in systems satisfying the free-end BCs. Finally, we conclude and discuss possible extensions in Sec. V.

\subsection{II. The SSH coupled-spring system with fixed-end boundary conditions}

Let's begin with the following 1D coupled-spring system, of which the Hamiltonian is given by
\bea
H_{\rm SSH}=\sum_{j=-\infty}^{\infty}\left\{  \frac{m}{2} \left[ \dot{a}_j^2(t) + \dot{b}_j^2(t) \right] + \frac{k_0}{2} \left[ a_j(t) - b_j (t) \right]^2 +  \frac{k_1}{2} \left[ a_{j+1}(t) - b_j(t)  \right]^2 \right\}.
\eea
Here, $j$ denotes the unit cell, and $k_0, k_1$ are the intra-cell and inter-cell spring constants, respectively.  $a_j(t) $ and $b_j(t)$ represent the distance from the equilibrium position. Because of its similarity to the SSH model, we will refer to it as the SSH coupled-spring system.  Using separation of variables $a_j(t) = {\rm e}^{-i \Om t} A_j$, $b_j(t) = {\rm e}^{-i \Om t} B_j$, we may achieve the following time-independent equation of motion (EOM):
\begin{subequations}
\label{Inf}
\begin{align}
\Omh^2  A_j + \left(\kh_0 B_j + \kh_1 B_{j-1}  \right) & = 0;  \label{Inf-A} \\
\Omh^2  B_j + \left(\kh_0 A_j + \kh_1 A_{j+1} \right) & = 0,  \label{Inf-B}
\end{align}
\end{subequations}
where $\kh_0 = k_0/m, \kh_1 = k_1/m$ and $\Omh^2 = \Om^2 - (\kh_0 + \kh_1)$.  It may be seen that the above equation is similar to that of the SSH model, and thus the results obtained for the SSH model may be carried over directly.

The simplest way to find the edge state is to consider a right semi-infinite chain of such a system:
\bea
H^{\rm R}_{\rm SSH}=\sum_{j=1}^{\infty}\left\{ \frac{m}{2} \left[ \dot{a}_j^2(t) + \dot{b}_j^2(t) \right] + \frac{k_0}{2} \left[ a_j(t) - b_j (t) \right]^2 +  \frac{k_1}{2} \left[ a_{j+1}(t) - b_j(t)  \right]^2  \right\} + \frac{k_1}{2} a_1^2(t).
\eea
Here, $a_1(t)$ describes the position of the site by the left edge of the system. Since there is now a left boundary, the EOM becomes
\begin{subequations}
\label{R-semi-inf}
\begin{align}
\Omh^2  A_j + \left(\kh_0 B_j + \kh_1 B_{j-1}  \right) & = 0,  \mbox{ for } j\ge 2; \label{R-semi-inf-A} \\
\Omh^2  B_j + \left(\kh_0 A_j + \kh_1 A_{j+1} \right) & = 0 ,  \mbox{ for } j\ge 1. \label{R-semi-inf-B}
\end{align}
\end{subequations}
Because of the additional spring with a force constant $k_1$ attached to $a_1(t)$, the system satisfies the fixed-end boundary condition (BC). Hence $A_1$ satisfies $\Omh^2 A_1 + \left(\kh_0 B_1  \right) = 0$, which may be simplified to
\bea
&\;& \hskip -3.1cm B_0 =0.
\eea
This is mathematically equivalent to the open BC in the SSH model. We may find the edge state of such a system by letting
\bea
\label{SSH-s-sol}
&\;& \hskip -3.1cm A_j =\a s^j, B_j=\b s^j,
\eea
with $|s|<1$. As a result, the EOM becomes
\begin{subequations}
\label{EOM SSH}
\begin{align}
\Omh^2 \a- \left(\kh_0 + \kh_1 s^{-1} \right)\b & =0, \label{EOM SSH-A} \\
\Omh^2 \b -\left(\kh_0 + \kh_1 s \right)\a  & =0. \label{EOM SSH-B}
\end{align}
\end{subequations}
The BC would require that $\b =0$. A non-trivial solution exists only if
\be
\label{edge-state}
\Omh  = 0\;\;  {\rm and }\;\; s= - k_0/k_1.
\ee
For the solution to be a physical state, we must have $k_1>k_0$. Thus, it will be a mid-gap state, with its energy lying exactly in the middle between the bottom of the upper band and the top of the lower band.  In the SSH model, the winding number $\n=1$ when this occurs, and we refer to it as the topological phase.

To confirm explicitly the BBC, we must consider a finite chain of the model and resort to numerically diagonalizing the matrix corresponding to the Hamiltonian of the system.  To be specific, let's focus on the case of a chain with $2N$ sites:
\bea
&\;& \hskip -4.5cm H^{\rm even}_{\rm SSH}=\sum_{j=1}^{N-1}\Biggl\{ \frac{m}{2} \left[ \dot{a}_j^2(t) + \dot{b}_j^2(t) \right] + \frac{k_0}{2} \left[ a_j(t) - b_j (t) \right]^2 +  \frac{k_1}{2} \left[ a_{j+1}(t) - b_j(t)  \right]^2  \Biggr\} \cr
&\;& \hskip -2.5cm  + \frac{m}{2} \left[ \dot{a}_N^2(t) + \dot{b}_N^2(t) \right] + \frac{k_0}{2} \left[ a_N(t) - b_N (t) \right]^2 + \frac{k_1}{2} \left[ a_1^2(t) +  b_N^2(t) \right].
\eea
The BCs become
\bea
&\;& \hskip -3.1cm B_0 =0, \quad A_{N+1}=0.
\eea
To find the solutions to the system, we again let $A_j =\a s^j,  B_j=\b s^j$ and it would lead to the equations shown in eq.~(\ref{EOM SSH}). Non-trivial solutions for $\a$ and $\b$ exist only if the determinant formed by the coefficients of $\a$ and $\b$ is vanishing, i.e.
\bea \label{Seculae eq 1}
&\;& \hskip -3.1cm \Omh^4 - \left\{\kh_0^2 + \kh_1^2 + \kh_0 \kh_1 \left(s +  s^{-1} \right) \right\} =0,
\eea
which we call the secular equation.  Since the above equation is quadratic in $s$ and the two solutions are reciprocal of each other, the most general solutions of $A_j$ and $B_j$ to the EOM are given by
\bea
&\;& \hskip -3.1cm A_j =\a_+ s^j +\a_-  s^{-j} , B_j=\b_+ s^j +\b_- s^{-j}.
\eea
Note that $\a_\pm$ and $\b_\pm$ are related through eq.~(\ref{EOM SSH}). By imposing the BCs and expressing $\a_\pm$ in terms of $\b_\pm$, we have
\begin{subequations}
\label{BC SSH}
\begin{align}
& \hskip 3.0cm \b_+ + \b_-   \hskip 2.28cm = 0, \label{BC SSH-A} \\
& \left(\kh_0 + \kh_1 s^{-1} \right) s^{N+1}\b_+  \left(\kh_0 + \kh_1 s \right)s^{-N-1} \b_-   = 0. \label{BC SSH-B}
\end{align}
\end{subequations}
Similarly, non-trivial solutions for $\b_+$ and $\b_-$ exist only if the determinant formed by their coefficients is vanishing. In terms of $u = \left(s+s^{-1}\right)/2$, the secular and characteristic equation of $s$ are
\begin{subequations}
\label{SSH-secular-characteristic}
\begin{align}
 & \hskip -1.0cm  \Omh^2 =  \pm \sqrt{ \kh_0^2 + \kh_1^2 + 2\kh_0 \kh_1 u }; \label{SSH-secular} \\
 & \hskip -1.0cm  k_1 U_{N-1} (u) + k_0 U_{N} (u) = 0. \label{SSH-secular-characteristic}
\end{align}
\end{subequations}
Here, $U_N(u)$ is the Chebysheve polynomial of the second kind of order $N$. Just like in the SSH model, edge states appear only if $k_1>k_0$. They are associated with the roots $s \approx -k_0/k_1, -k_1/k_0$, which may be explicitly verified by numerical calculation. There are two edge states since there are now two boundaries.  Because of the tunneling effect, there is mixing between the edge states which leads to exponentially small energy splitting.

\subsection{III. The SSH coupled-spring system with free-end boundary conditions}

For the SSH coupled-spring systems, another type of BC also exists, the free-end BC. Let's now consider a finite chain with $2N$ sites:
\bea
&\;& \hskip -4.5cm H^{\rm R}_{\rm SSH}=\sum_{j=1}^{N-1}\Biggl\{ \frac{m}{2} \left[ \dot{b}_j^2(t) + \dot{b}_j^2(t) \right] + \frac{k_0}{2} \left[ a_j(t) - b_j (t) \right]^2 +  \frac{k_1}{2} \left[ a_{j+1}(t) - b_j(t)  \right]^2  \Biggr\} \cr
&\;& \hskip -2.5cm  + \frac{m}{2} \left[ \dot{a}_N^2(t) + \dot{b}_N^2(t) \right] + \frac{k_0}{2} \left[ a_N(t) - b_N (t) \right]^2.
\eea
After simplification, the BCs take the following form
\begin{subequations}
\begin{align}
& A_1 - B_0 \quad\;\; = 0, \\
& A_{N+1} - B_N = 0.
\end{align}
\end{subequations}
By introducing $r_\pm = \a_\pm/\b_\pm,$ the above BCs become
\begin{subequations}
\label{BC SSH-free-even}
\begin{align}
& \left( r_+ s -1 \right)\b_+ + \left( r_- s^{-1} -1 \right)\b_- \hskip 1.03cm = 0
\label{BC SSH-free-even-A}\\
& \left( r_+ s -1 \right)s^N \b_+ + \left( r_- s^{-1} -1 \right) s^{-N}\b_- = 0.
\label{BC SSH-free-even-B}
\end{align}
\end{subequations}

It may be checked that $r_+ r_- = 1$.  This gives rise to the condition that $s^{2N} = 1$, and hence $s=e^{i\pi l/N}$. For $k_0>k_1$, the normal mode frequencies are  given by
\bea
&\;& \hskip -2.1cm \Omh^2 =
\begin{cases}
\;\;\; \sqrt{ \kh_0^2 + \kh_1^2 + 2 \kh_0 \kh_1 \cos\left[\pi l/(N+1)\right] },  & l = 1, \ldots, N; \\
-\sqrt{ \kh_0^2 + \kh_1^2 + 2 \kh_0 \kh_1 \cos\left[\pi l/(N+1)\right] }, & l = 0, \ldots, N-1.
\end{cases}
\eea
In contrast, for For $k_0<k_1$
\bea
&\;& \hskip -2.1cm \Omh^2 =
\begin{cases}
\;\;\; \sqrt{ \kh_0^2 + \kh_1^2 + 2 \kh_0 \kh_1 \cos\left[\pi l/(N+1)\right] },  & l = 1, \ldots, N-1; \\
-\sqrt{ \kh_0^2 + \kh_1^2 + 2 \kh_0 \kh_1 \cos\left[\pi l/(N+1)\right] }, & l = 0, \ldots, N.
\end{cases}
\eea

On the other hand, when there are $2N+1$ sites in the system, we have
\bea
&\;& \hskip -3.5cm H^{\rm R}_{\rm SSH}=\sum_{j=1}^{N}\Biggl\{ \frac{m}{2} \left[ \dot{a}_j^2(t) + \dot{b}_j^2(t) \right] + \frac{k_0}{2} \left[ a_j(t) - b_j (t) \right]^2 +  \frac{k_1}{2} \left[ a_{j+1}(t) - b_j(t)  \right]^2  \Biggr\} + \frac{m}{2} \dot{a}_{N+1}^2(t).
\eea
The BCs now become
\begin{subequations}
\begin{align}
& A_1 - B_0 \quad\quad\;\; = 0, \\
& B_{N+1} - A_{N+1} = 0.
\end{align}
\end{subequations}
Similarly, we have
\begin{subequations}
\label{BC SSH-free-odd}
\begin{align}
& \left( r_+ s -1 \right)\b_+ + \left( r_- s^{-1} -1 \right)\b_- \hskip 1.03cm = 0
\label{BC SSH-free-odd-A}\\
& \left( r_+ -1 \right)s^{N+1} \b_+  + \left( r_-  -1 \right) s^{-N-1}\b_- = 0.
\label{BC SSH-free-odd-B}
\end{align}
\end{subequations}
Again, the recursion relation may be solved analytically and $s=e^{i2\pi l/(2N+1)}$. In this case, the normal mode frequencies are  given by
\bea
&\;& \hskip -2.1cm \Omh^2 =
\begin{cases}
\;\;\; \sqrt{ \kh_0^2 + \kh_1^2 + 2 \kh_0 \kh_1 \cos\left[2\pi l/(2N+1)\right] },  & l = 1, \ldots, N; \\
-\sqrt{ \kh_0^2 + \kh_1^2 + 2 \kh_0 \kh_1 \cos\left[2\pi l/(2N+1)\right] }, & l = 0, \ldots, N,
\end{cases}
\eea
independent of whether $k_0>k_1$ or $k_0<k_1$. The most surprising result here is that for the free-end BC, there is no mid-gap edge state even if the system is in the "topological phase" ($k_1>k_0$). This gives another explicit example that BBC depends not only on the values of physical parameters but also on the BCs. If the BCs are fixed-end and the system is in the topological phase, then edge states would appear on the boundary.  On the other hand, when the BCs are free-end, then there would be no edge state even if the system is in the topological phase. Some may argue that in the case of free-end BCs, the ``chiral symmetry'' is violated on the boundary and that this is why there is no edge state in the topological phase. However, a counter-example is given by the case that the BCs are fixed-end on one of the boundaries and free-end on the other. As it turns out, the boundary with the fixed-end BC would still admit an edge state if the system is in the topological phase.

To be more specific, we choose the left and right BCs to be fixed-end and free-end, respectively.  When there are $2N$ sites, the characteristic equation is
\bea
&\;& \hskip -3.1cm \kh_0 U_N(u)
+ \left( \kh_1\pm \sqrt{ \kh_0^2 + \kh_1^2 + \kh_1 + 2\kh_0\kh_1 u  } \right) U_{N-1}(u) = 0,
\eea
When there are $2N+1$ sites, the characteristic equation becomes
\bea
&\;& \hskip -3.1cm \left(\kh_0  \pm \sqrt{ \kh_0^2 + \kh_1^2 + \kh_0 +  2\kh_0\kh_1 u  } \right) U_N(u) +  \kh_1 U_{N-1}(u) = 0.
\eea
Here, the plus and minus signs refer to the upper and lower bands.

\begin{figure}[hbt!]
\centering
\subfloat[]{\includegraphics[width=0.40\textwidth]{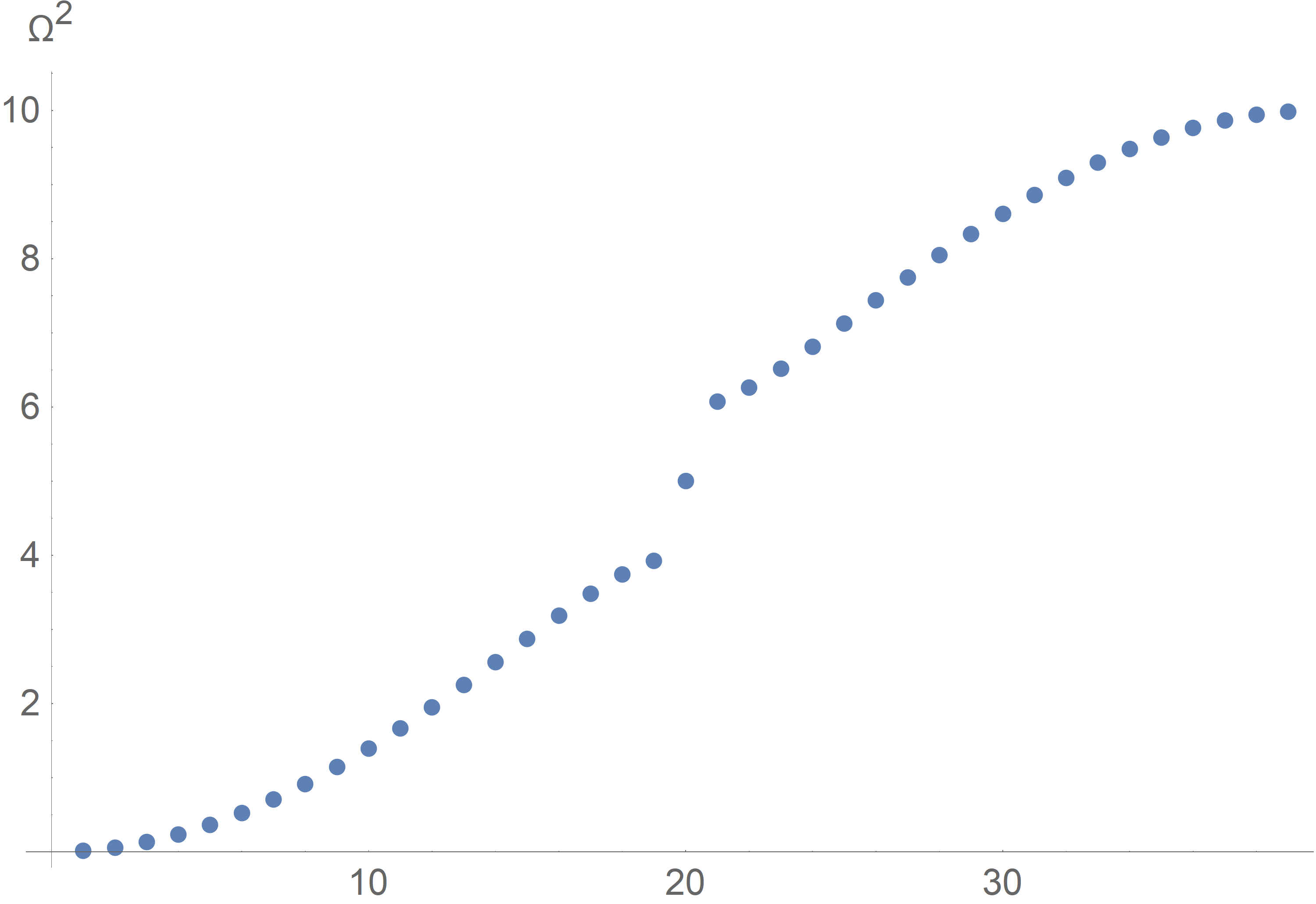}}\hskip 0.5cm
\subfloat[]{\includegraphics[width=0.40\textwidth]{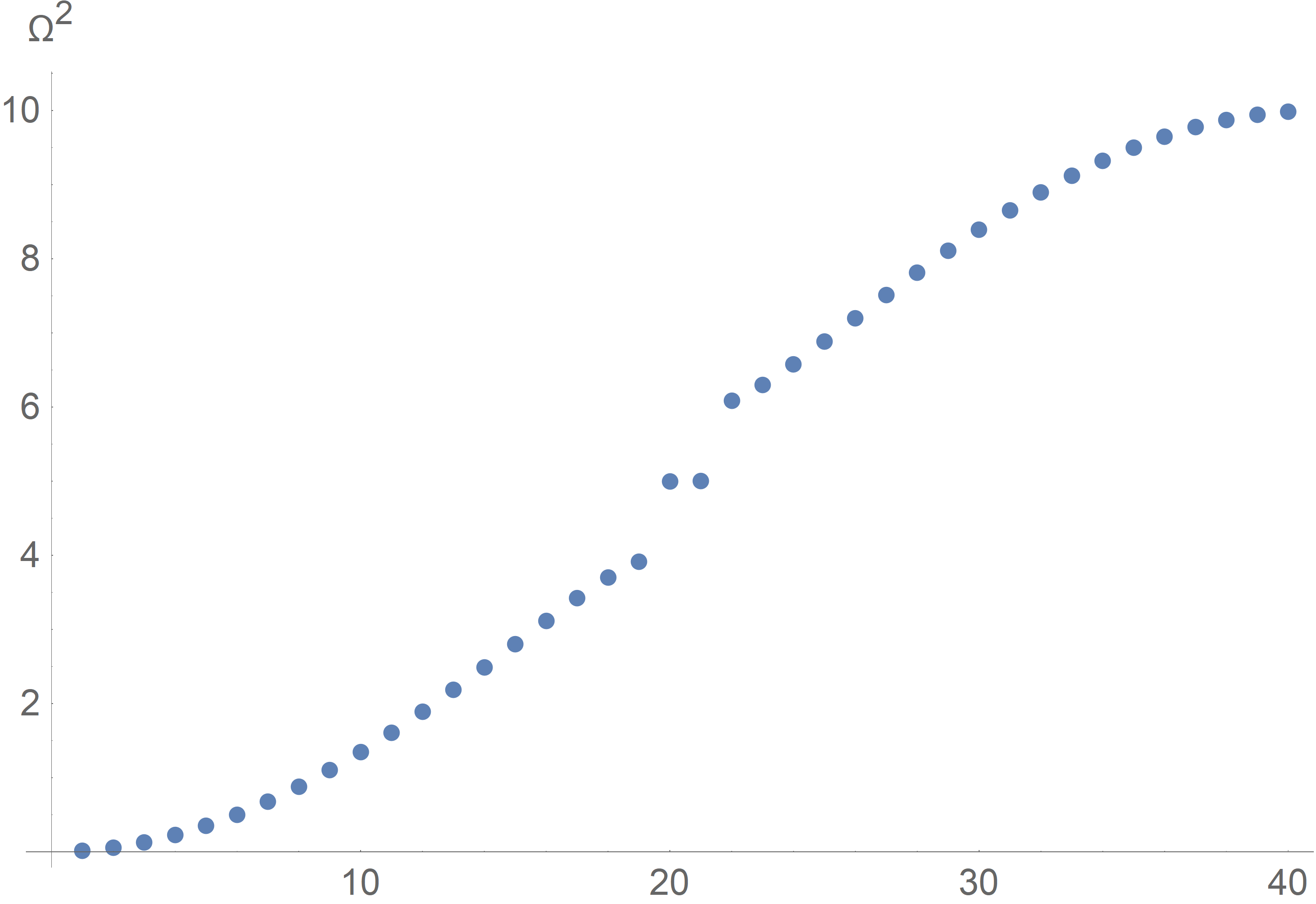}}\\
\subfloat[]{\includegraphics[width=0.40\textwidth]{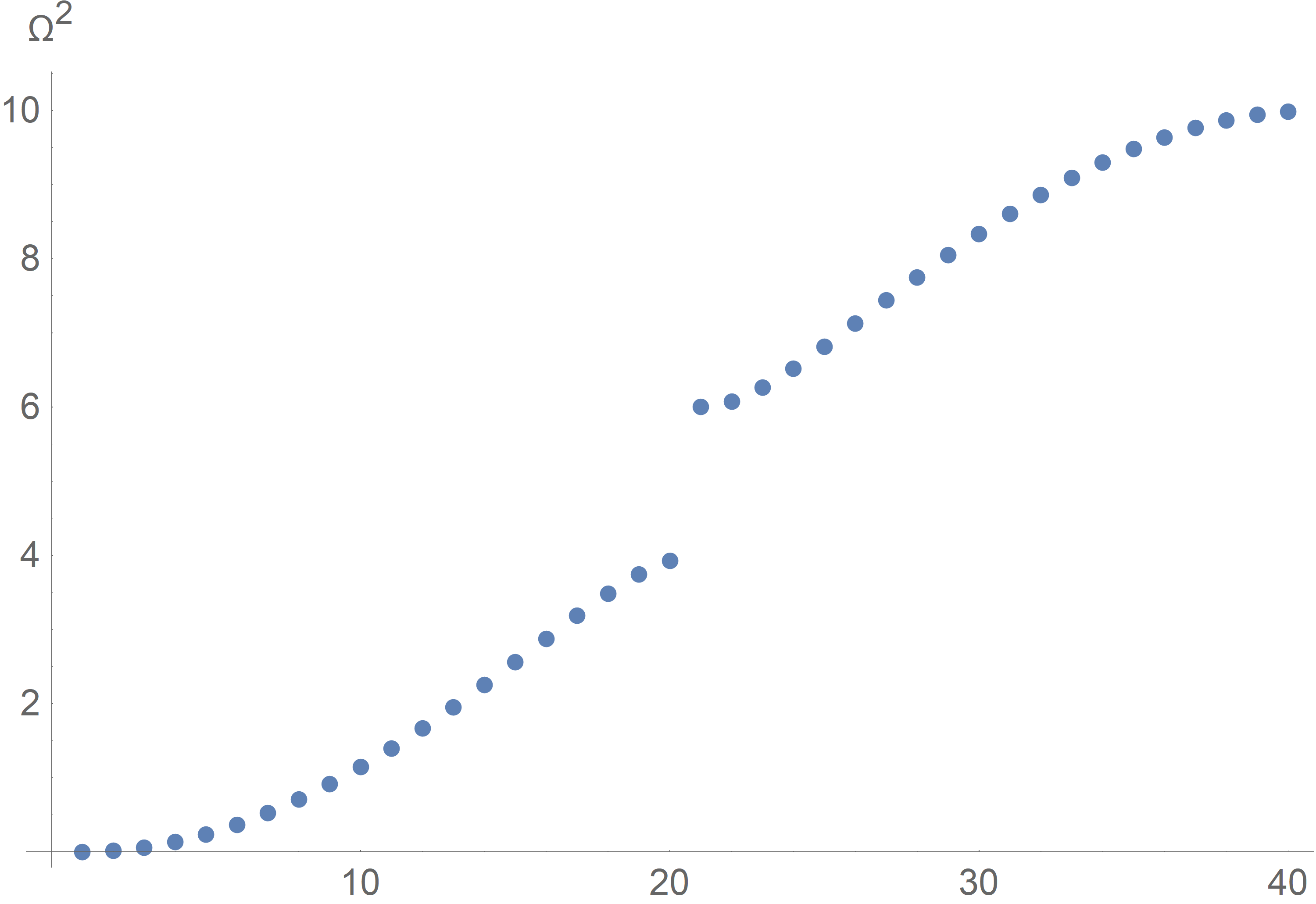}}\hskip 0.5cm
\subfloat[]{\includegraphics[width=0.40\textwidth]{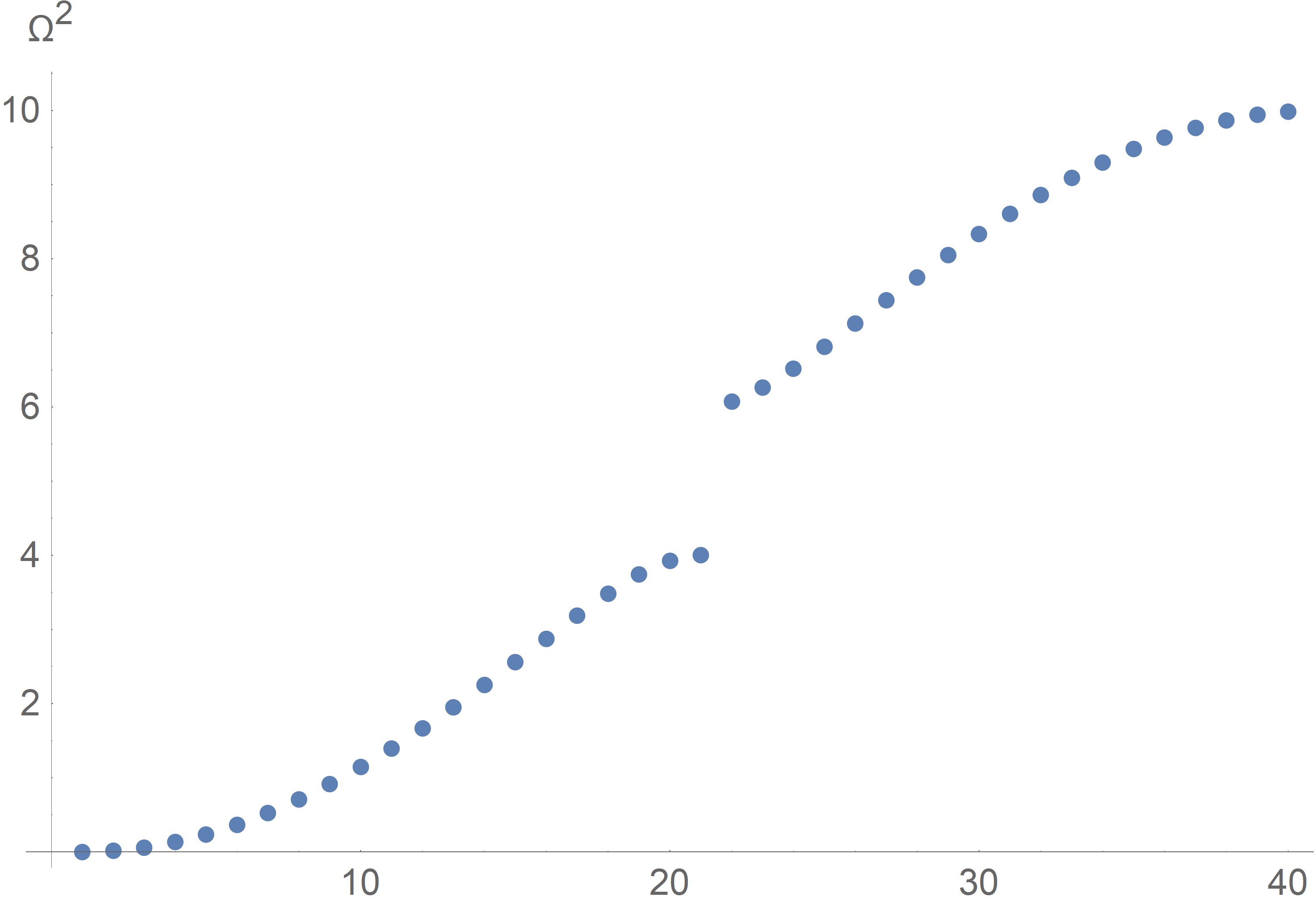}}\\
\subfloat[]{\includegraphics[width=0.40\textwidth]{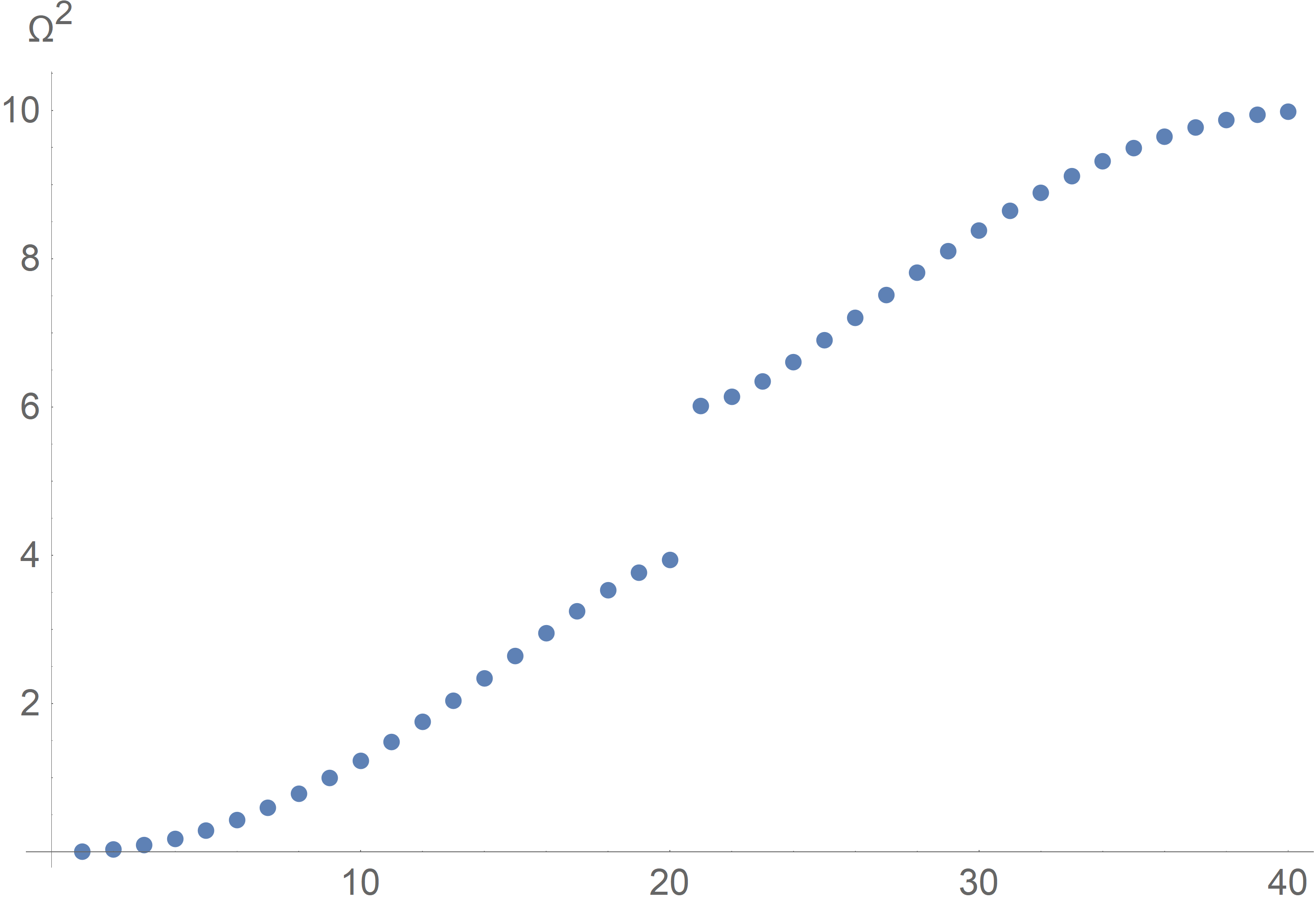}}\hskip 0.5cm
\subfloat[]{\includegraphics[width=0.40\textwidth]{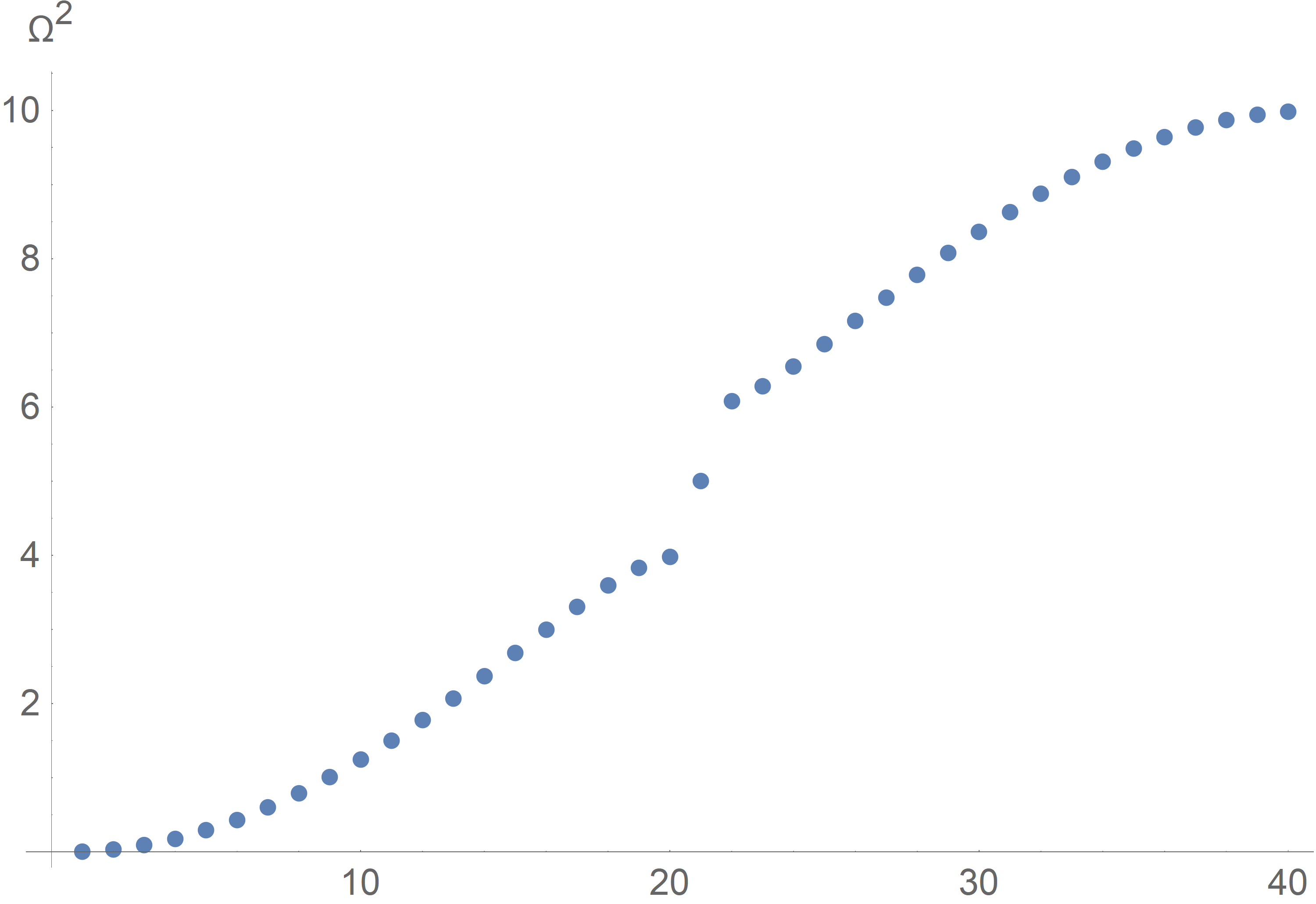}}\\
\caption{The energy spectrum of the SSH coupled-spring systems. The total number of sites $N_{\rm tot} = 40$. The parameters are chosen to be $\left( \kh_0, \kh_1 \right) = (3, 2)$ and  $\left( \kh_0, \kh_1 \right) = (2, 3)$ in (a), (c) (e) and (d), (d), (f), respectively. (a), (b) The BCs are fixed-end both on the left and right boundaries, with (a) and (b) in the trivial and topological phases, respectively.  There are two mid-gap edge states in (b) consistent with the BBC. (c), (d) The BCs are free-end both on the left and right boundaries, with (c) and (d) in the trivial and topological phases, respectively.  Note that there is no mid-gap edge state in either case. (e), (f) The BCs are fixed-end and free-end on the left and right boundaries, with (e) and (f) in the trivial and topological phases, respectively.  Note that there is only one mid-gap edge state in (f). Thus, the BBC only holds on the left boundary, which satisfies the fixed-end BC.}  \label{fig1 SSH-spring}
\end{figure}

\begin{figure}[hbt!]
\centering
\subfloat[]{\includegraphics[width=0.40\textwidth]{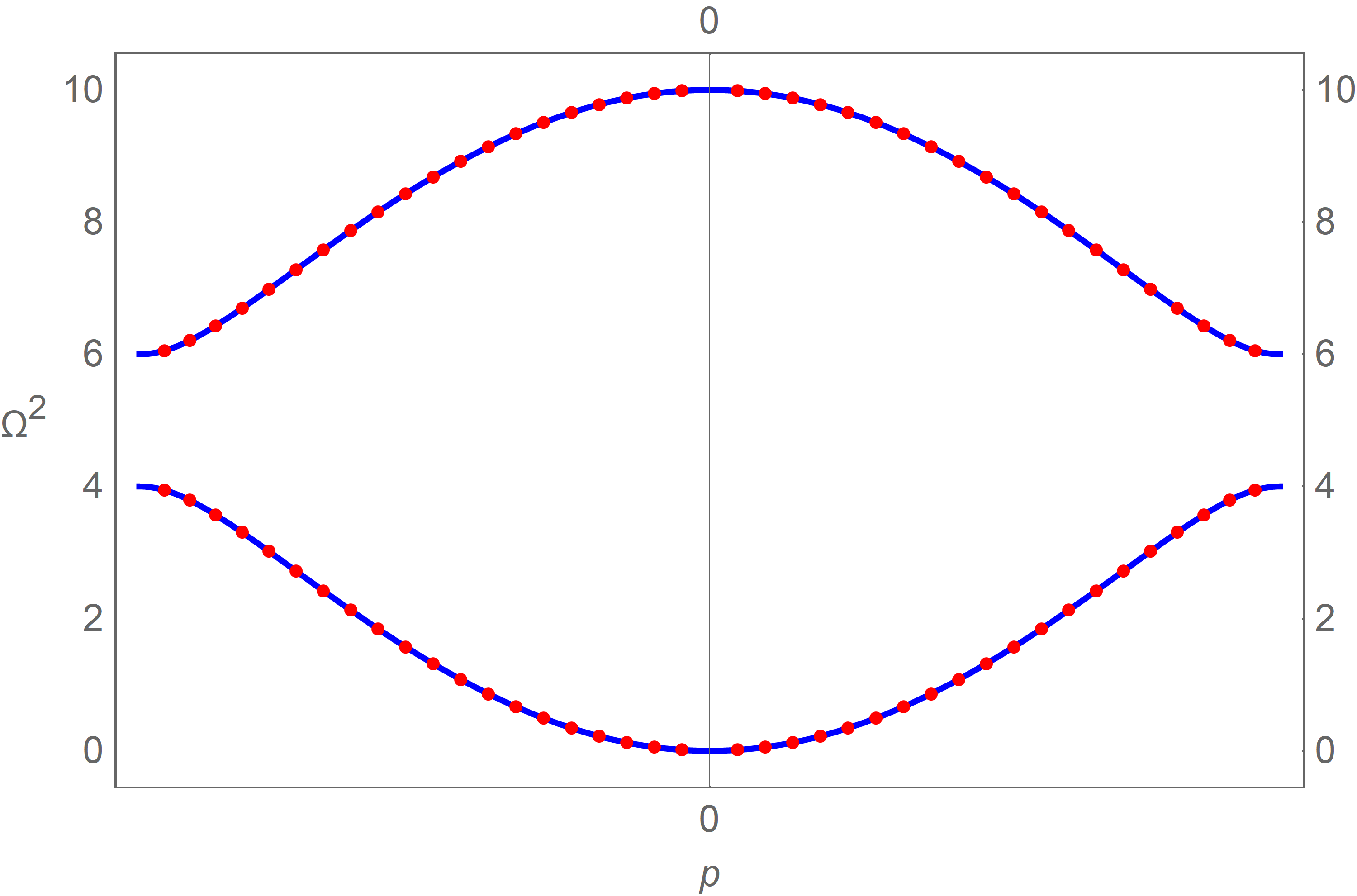}}\hskip 0.5cm
\subfloat[]{\includegraphics[width=0.40\textwidth]{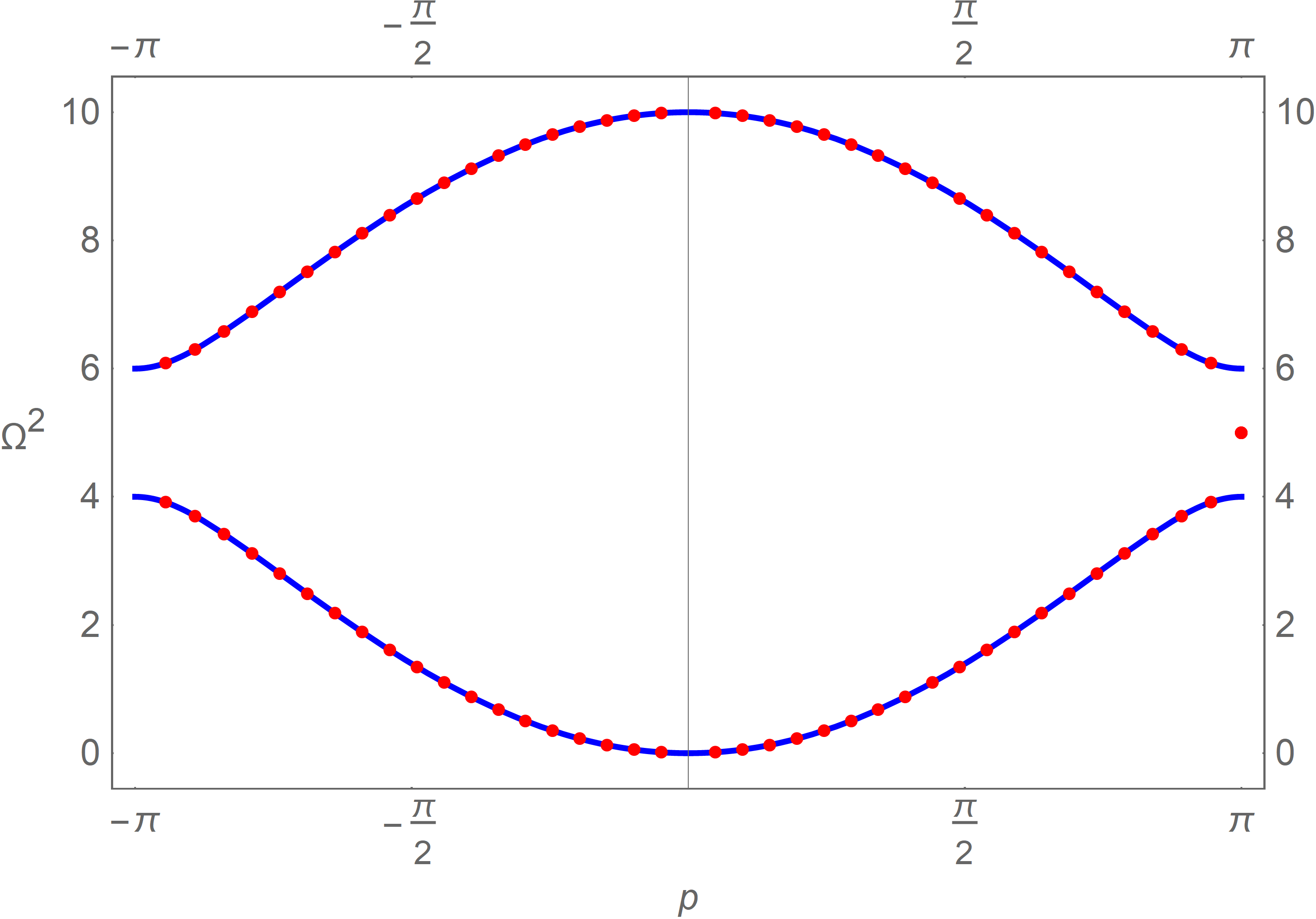}}\\
\subfloat[]{\includegraphics[width=0.40\textwidth]{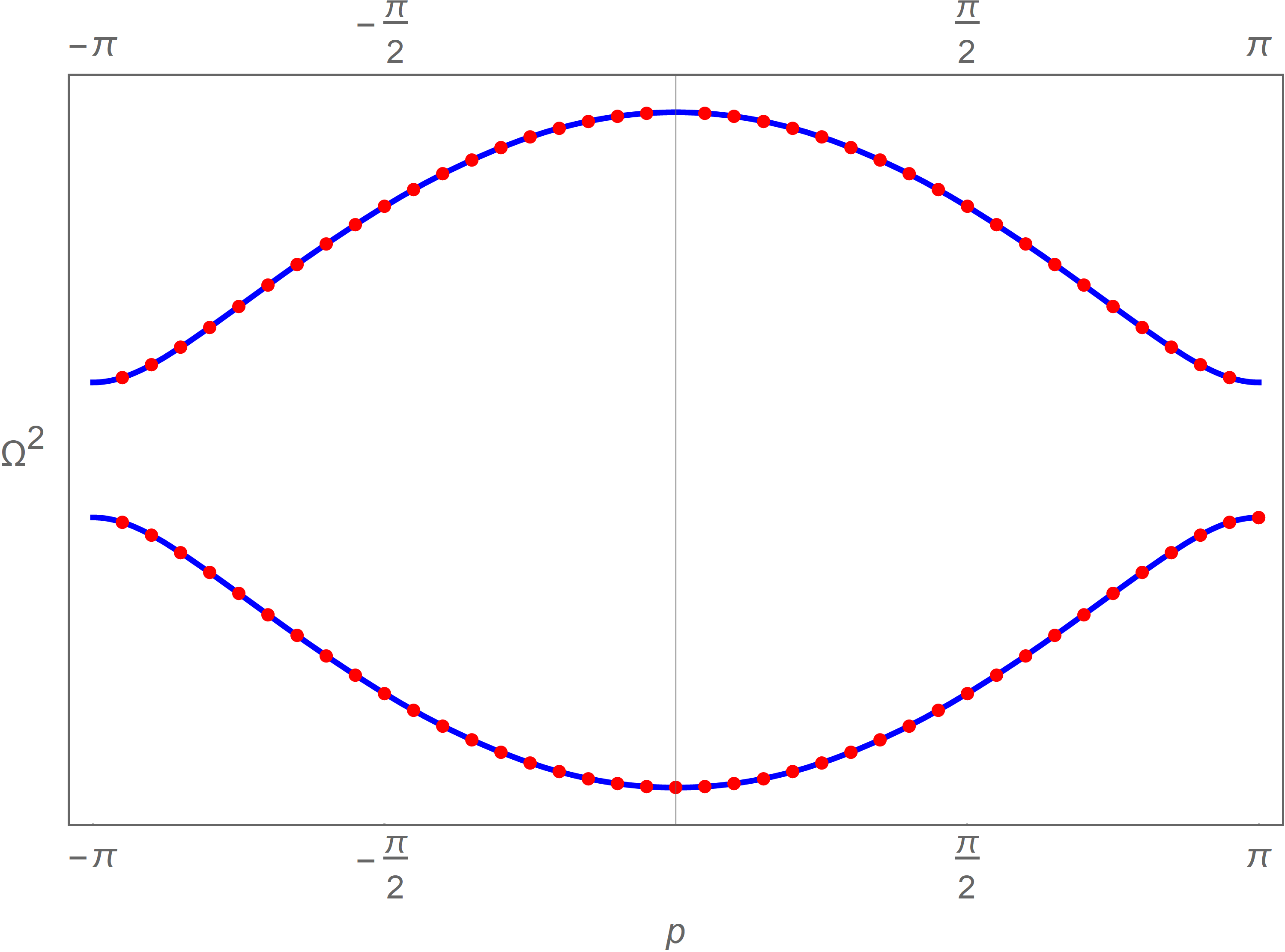}}\hskip 0.5cm
\subfloat[]{\includegraphics[width=0.40\textwidth]{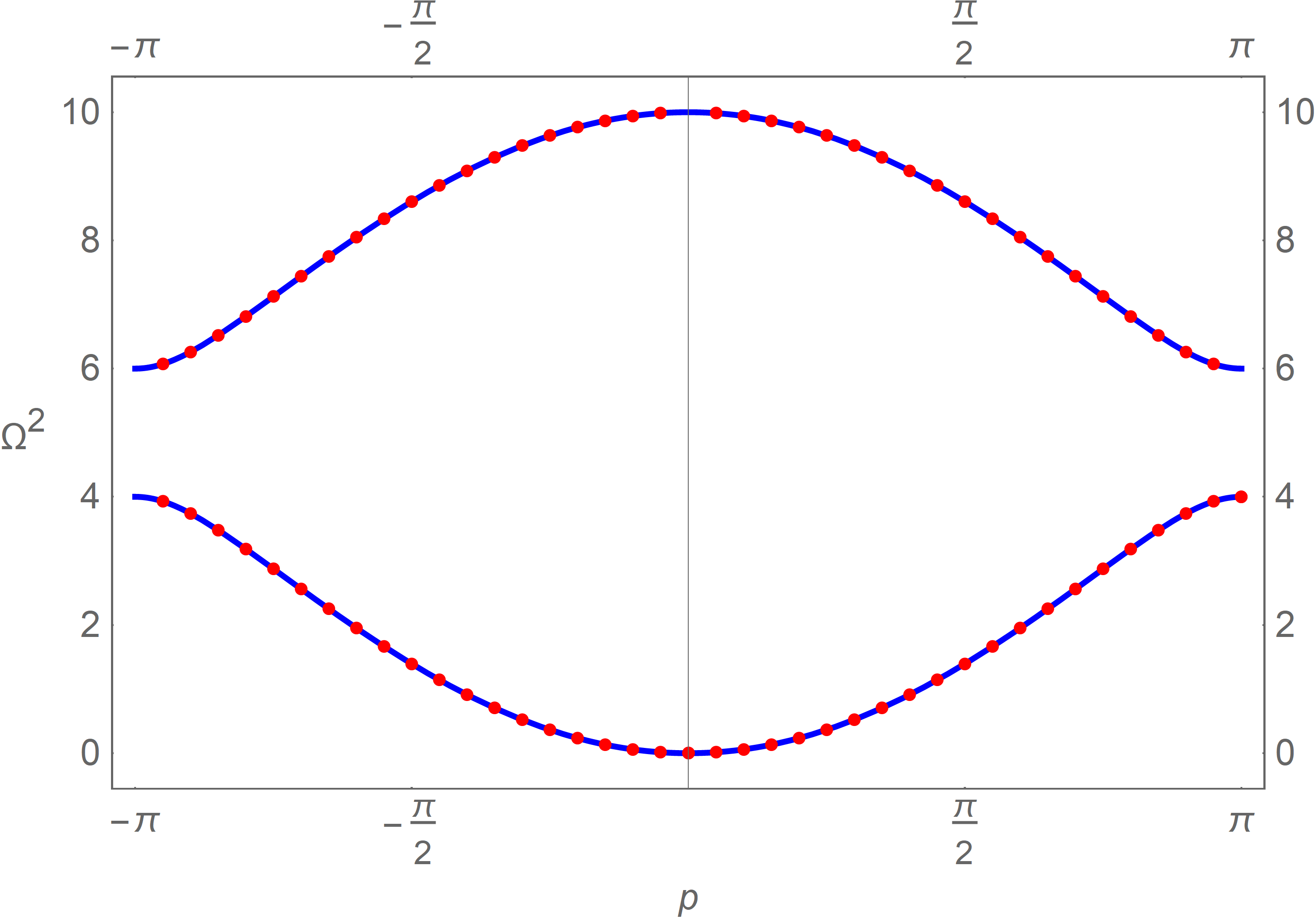}}\\
\subfloat[]{\includegraphics[width=0.40\textwidth]{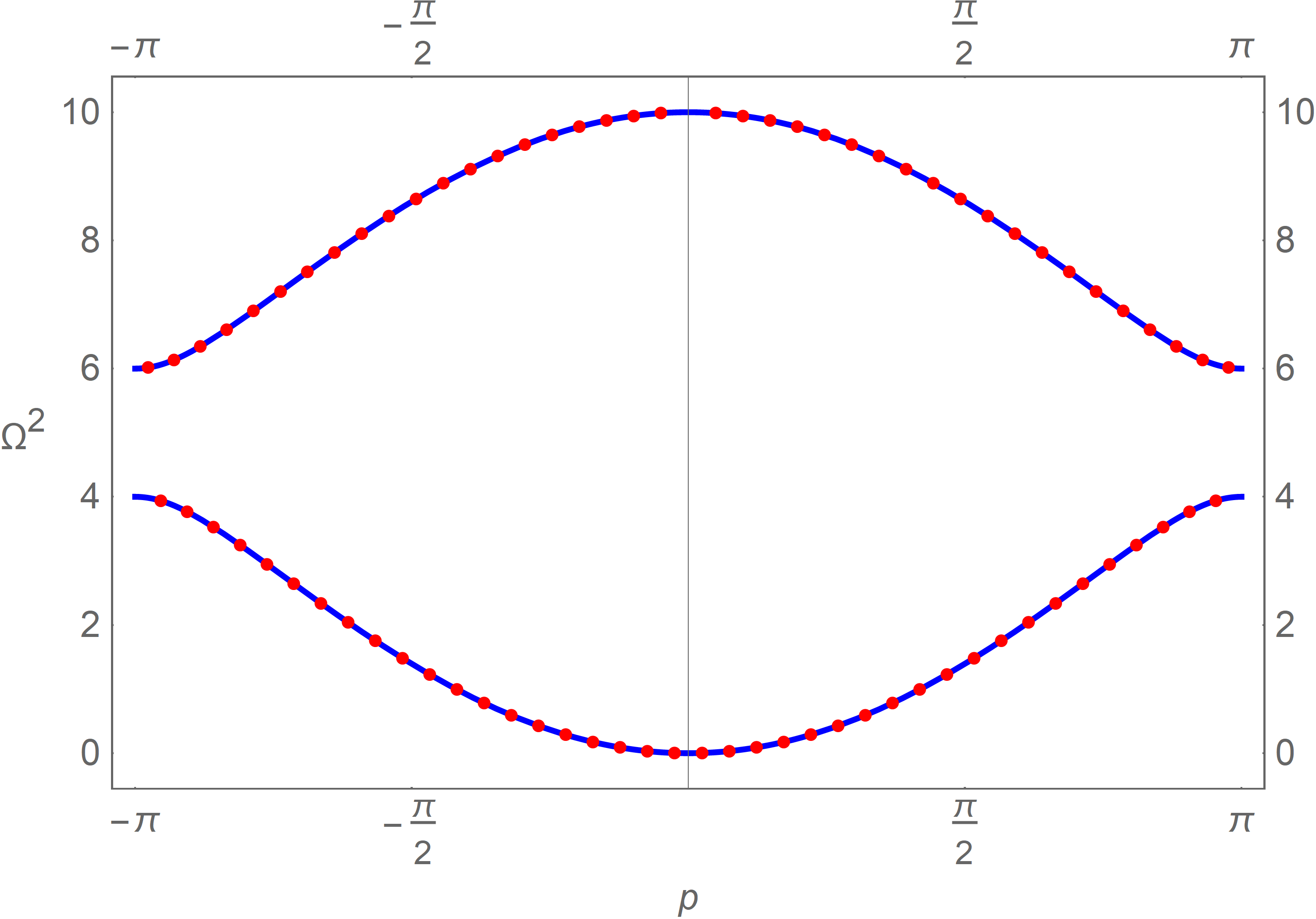}}\hskip 0.5cm
\subfloat[]{\includegraphics[width=0.40\textwidth]{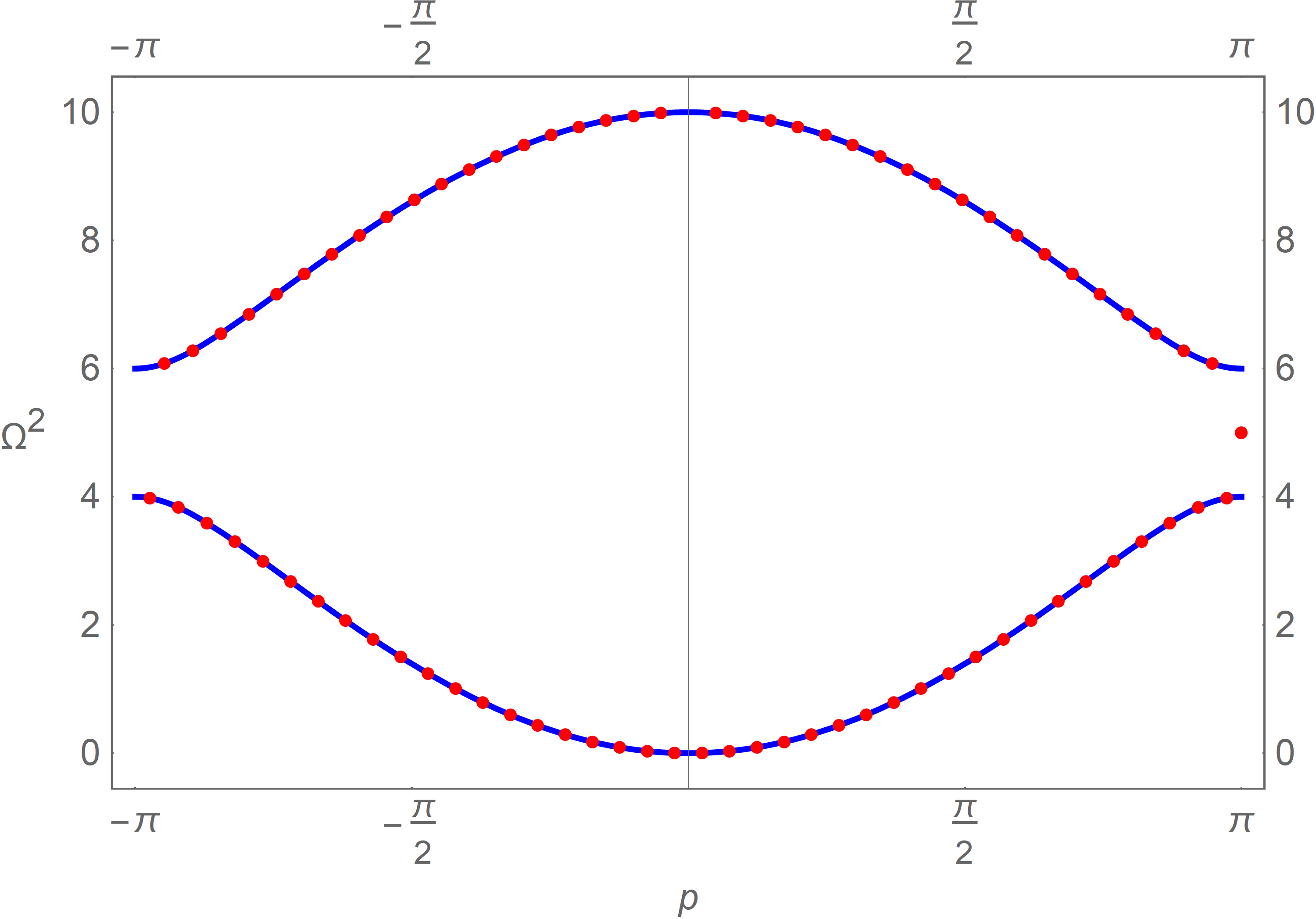}}\\
\caption{The energy spectrum of the SSH coupled-spring systems. The total number of sites $N_{\rm tot} = 40$. The parameters are chosen to be $\left( \kh_0, \kh_1 \right) = (3, 2)$ and  $\left( \kh_0, \kh_1 \right) = (2, 3)$ in (a), (c) (e) and (d), (d), (f), respectively. (a), (b) The BCs are fixed-end both on the left and right boundaries, with (a) and (b) in the trivial and topological phases, respectively.  There are two mid-gap edge states in (b), but they are almost on top of each other. (c), (d) The BCs are free-end both on the left and right boundaries, with (c) and (d) in the trivial and topological phases, respectively.  (e), (f) The BCs are fixed-end and free-end on the left and right boundaries, respectively. (e) is in the trivial phase and (f) is in the topological phase.  In (f), there is only one mid-gap edge state, which exists on the left boundary.}  \label{fig2 SSH-spring}
\end{figure}

\begin{figure}[hbt!]
\centering
\subfloat[]{\includegraphics[width=0.40\textwidth]{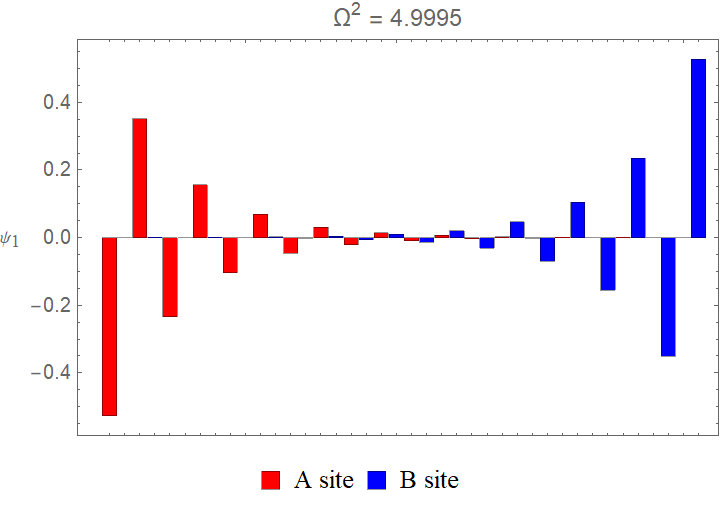}}\hskip 0.5cm
\subfloat[]{\includegraphics[width=0.40\textwidth]{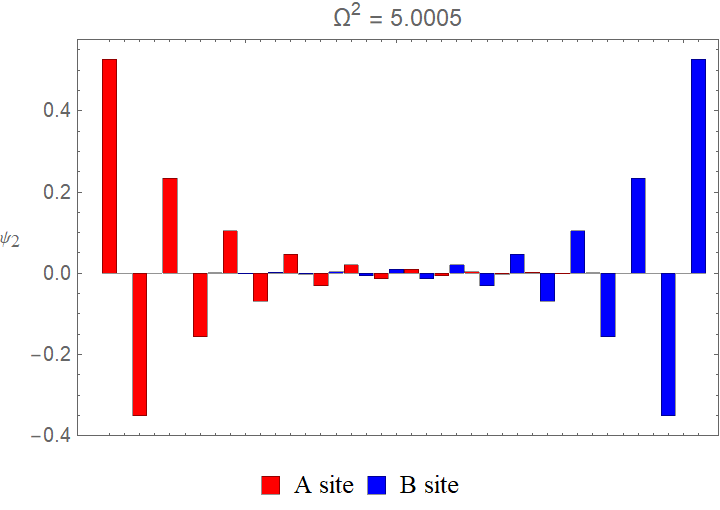}}\\
\subfloat[]{\includegraphics[width=0.40\textwidth]{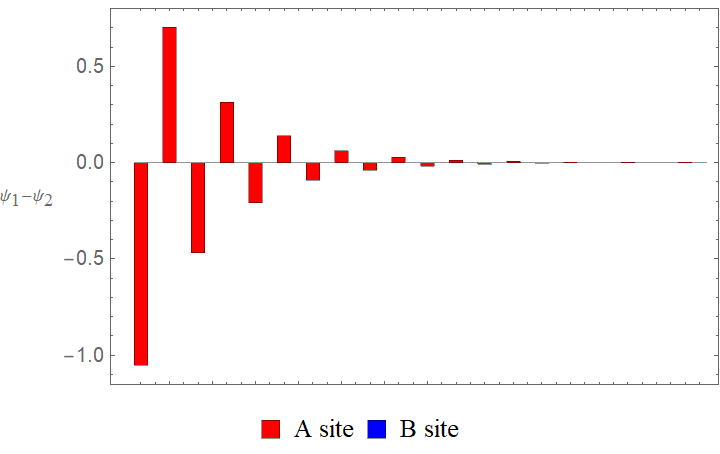}}\hskip 0.5cm
\subfloat[]{\includegraphics[width=0.40\textwidth]{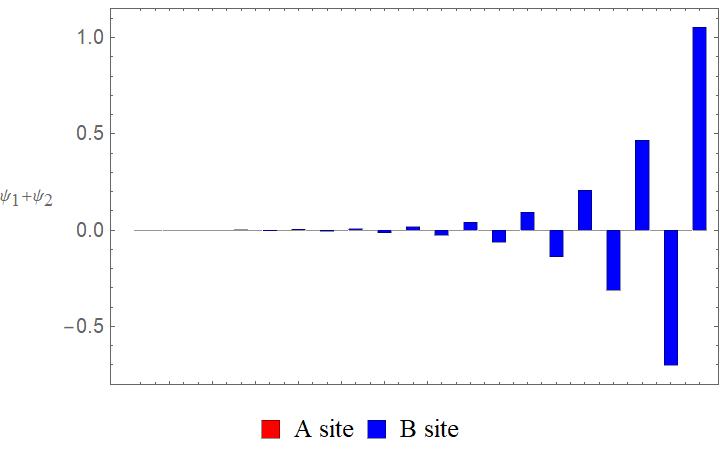}}\\
\subfloat[]{\includegraphics[width=0.40\textwidth]{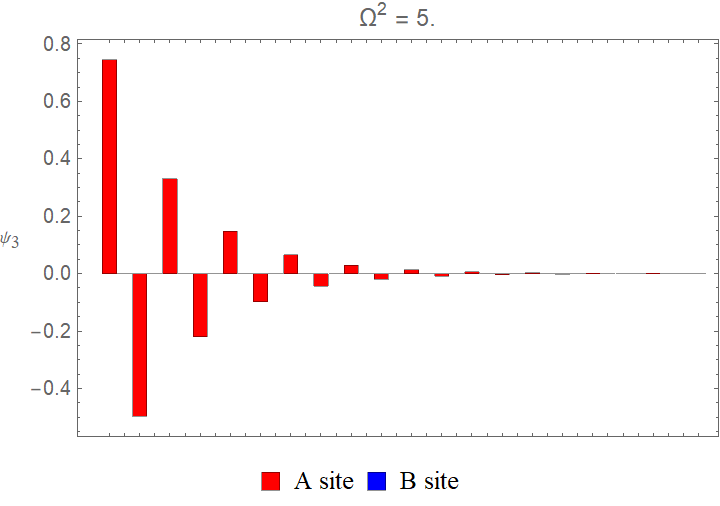}}\hskip 0.5cm
\caption{Wave functions of the mid-gap edge states. The total number of sites $N_{\rm tot} = 40$ and $\left( \kh_0, \kh_1 \right) = (2, 3)$  (a), (b) The BCs are fixed-end on both boundaries. Because of ``quantum'' tunneling, the normal modes are mixing of the left and right edge states. (c), (d) The difference and sum of the two edge states. The former and latter ones are left and right edge states, respectively.  (e) The BCs are fixed-end and free-end on the left and right boundaries, respectively. Note there is only one mid-gap edge state which is a pure left edge state in contrast to the case with fixed-end BCs on both boundaries.}  \label{fig3 SSH-spring}
\end{figure}

All these predictions may be confirmed numerically. The results are shown in Fig.~\ref{fig1 SSH-spring} to \ref{fig3 SSH-spring}. Although we have checked both the cases of even and odd numbers of sites to ensure the correctness of our results, we only show those with an even number of sites. In Fig.~\ref{fig1 SSH-spring}, we show the energy spectra for various BCs. The total number of sites  $N_{\rm tot} = 40$. The parameters are chosen to be $\left( k_0, k_1 \right) = (3, 2)$ and $\left( k_0, k_1 \right) = (2, 3)$ for the trivial topological phases, respectively. Three different kinds of BCs are considered: (i) fixed-end on both boundaries, (ii) free-end on both boundaries, (iii) fixed-end on the left boundary, and free-end on the right boundary.

By solving the corresponding characteristic equations, we can also obtain the spectra of the systems. We check their validity by comparing the spectra obtained by solving the characteristic equations with those achieved from numerical diagonalization. From these results, we can make plots of $\Om^2$ versus lattice momentum $p$, and the results are shown in Fig.~\ref{fig2 SSH-spring}.

The wave functions of the mid-gap edge states are shown in Fig.~\ref{fig3 SSH-spring}. When the BCs are fixed-end on both boundaries, there would be an edge state on each boundary. Because of tunneling, the normal modes are mixing of the left and right edge states. By taking the difference and sum of the two edge states, we see they are indeed the left and right edge states. When the BCs are fixed-end and free-end on the left and right boundaries, there is only one mid-gap edge state. It is a pure left edge state in contrast to the case with fixed-end BCs on both boundaries.

\subsection{IV. The SSH coupled-spring system with general boundary conditions}

As the BCs of a system have a strong influence on the existence of edge states, it is desirable to consider a more general BC and study in detail how it affects the edge states.  To simplify the investigation, let's consider a right semi-infinite chain of SSH coupled-spring system satisfying the fixed-end BC, with the force constant of the spring attached to $a_1(t)$ from the left changed to $K$. The corresponding Hamiltonian is
\bea
&\;& \hskip -3.5cm H^{\rm R}_{{\rm SSH-}K-1}=\sum_{j=1}^{\infty}\Biggl\{ \frac{m}{2} \left[ \dot{a}_j^2(t) + \dot{b}_j^2(t) \right] + \frac{k_0}{2} \left[ a_j(t) - b_j (t) \right]^2 +  \frac{k_1}{2} \left[ a_{j+1}(t) - b_j(t)  \right]^2  \Biggr\} + \frac{K}{2} a_1(t) ^2,
\eea
When $K=0$, it reduces to the free-end BC. On the other hand, when $K = k_1$, it becomes the usual fixed-end BC that we considered in Section II.  When $K$ varies in the range $[0, k_1]$, the BC interpolates between the two limits. Note that as $K\to \infty$, it again approaches the fixed-end BC, where the boundary site is at $b_1(t)$ instead of $a_1(t)$.

The BC for $a_1$ and $b_1$ are given by
\begin{subequations}
\label{BC SSH-fix-R-inf}
\begin{align}
& \Om^2 A_1 + \Kh ( - A_1) + \kh_0 (B_1 - A_1) \hskip 0.60 cm   = 0.
\label{BC SSH-fix-R-inf-A1} \\
& \Om^2 B_1 + \kh_0 (A_1 - B_1) + \kh_1 (A_2 - B_1)  = 0.
\label{BC SSH-fix-R-inf-B1}
\end{align}
\end{subequations}
The edge states in this system should be depicted by the form given in eq.~(\ref{SSH-s-sol}) with $|s|<1$. The secular equation is still given by eq.~(\ref{SSH-secular}). The corresponding characteristic equation of $s$ may be obtained from the condition that the determinant formed by the coefficients of $\a$ and $\b$ is vanishing:
\bea
&\;& \hskip -4.5cm \left( \Kh - \kh_1 \right) A_1 + \kh_1 B_0 = 0 \Rightarrow \sqrt{\kh_0 + \kh_1 s} \left\{ \kh_1 \sqrt{\kh_0 + \kh_1 s} + \left (\Kh - \kh_1 \right) s \sqrt{\kh_0  + \kh_1 s^{-1}} \right\}=0.
\eea
After some algebra, the equation may be simplified to
\bea
&\;& \hskip -4.5cm   f_1(s) = \left (\Kh - \kh_1 \right)^2 s \left( \kh_0 s + \kh_1 \right) - \kh_1^2 \left( \kh_0 + \kh_1 s \right)  = 0,
\eea
with its roots given by
\bea
\label{Fixed-end-Edge-s}
&\;& \hskip -4.5cm   s_{\pm} = \frac{k_1 K(2k_1-K) \pm  k_1 \sqrt{4k_0^2 (k_1 -K)^2 + K^2 (2k_1 - K)^2}}{2k_0 (k_1- K)^2 }.
\eea
It is easy to see that both solutions are real, and they are of opposite signs.

Since
\bea
&\;& \hskip -6.5cm
\begin{cases}
f_1(-1)  = & \Kh(\kh_1 - \kh_0) (2 \kh_1 - \Kh ), \\
f_1(0) \;\;\;  = & -\kh_0 \kh_1^2, \\
f_1(1)\;\;\;  = & -\Kh(\kh_1 + \kh_0) (2 \kh_1 - \Kh ).
\end{cases}
\eea
From the signs of $f_1(-1), f_1(0)$, and $f_1(1)$, we will be able to tell how many roots have a magnitude less than one. Let's summarize the results here:
\bee[label=\arabic*)]
\item $\kh_0>\kh_1$, the trivial phase:
\bee[label=\roman*)]
\item $\Kh \le  2\kh_1$. No edge state.

\item $\Kh>2 \kh_1$. Two edge states. Their energies are near the top of the lower band and above the top of the upper band, respectively.
\eee

\item $\kh_0<\kh_1$, the topological phase:
\bee[label=\roman*)]
\item $\Kh \le  2\kh_1$. One mid-gap edge state with energy between the top of the lower band and the bottom of the upper band.

\item $\Kh>2 \kh_1$. One edge state with energy above the upper band.
\eee

\eee
In the topological phase, the energy of the mid-gap edge state decreases with $K$ and approaches the top of the lower band. When $K= 0$, we expect the edge state will disappear, since the system reduces to the one with a free-end BC in this limit.

Again, numerical check may be carried out using a finite chain. To be specific, we require the right boundary to satisfy the usual fixed-end BC. For $N_{\rm tot} = 2N$, it may be shown after some algebra that the characteristic equation of $s$ for such a system is given by
\bea
&\;& \hskip -3.1cm  \pm  \left( \kh_1 - \Kh \right) \sqrt{ \kh_0^2 + \kh_1^2 + 2\kh_0 \kh_1 u }\, U_{N-1}(u)
+ \kh_1 \left\{ \kh_0 U_N(u) + \kh_1 U_{N-1}(u)  \right\} = 0,
\eea
where $\Kh = K/m$.  Again, the plus and minus signs indicate the upper and lower bands.  If there are $2N+1$ sites, we have instead
\bea
&\;& \hskip -3.1cm  \pm  \kh_1 \sqrt{ \kh_0^2 + \kh_1^2 + 2\kh_0 \kh_1 u }\, U_{N}(u)
+ \left(\kh_1 -\Kh \right) \left\{ \kh_1 U_{N}(u) + \kh_0 U_{N-1}(u)  \right\} = 0.
\eea

\begin{figure}[hbt!]
\centering
\subfloat[]{\includegraphics[width=0.40\textwidth]{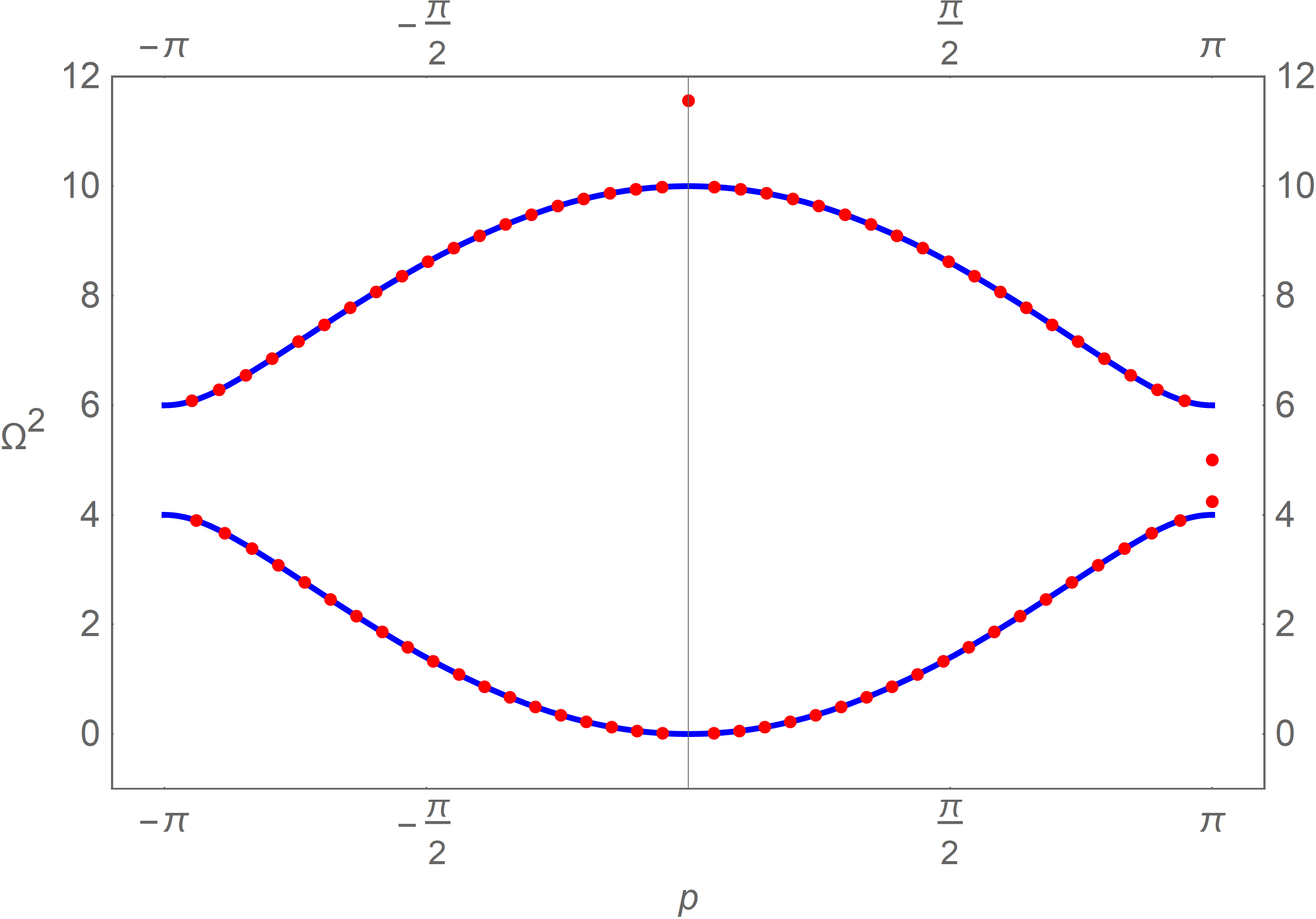}}\hskip 0.5cm
\subfloat[]{\includegraphics[width=0.40\textwidth]{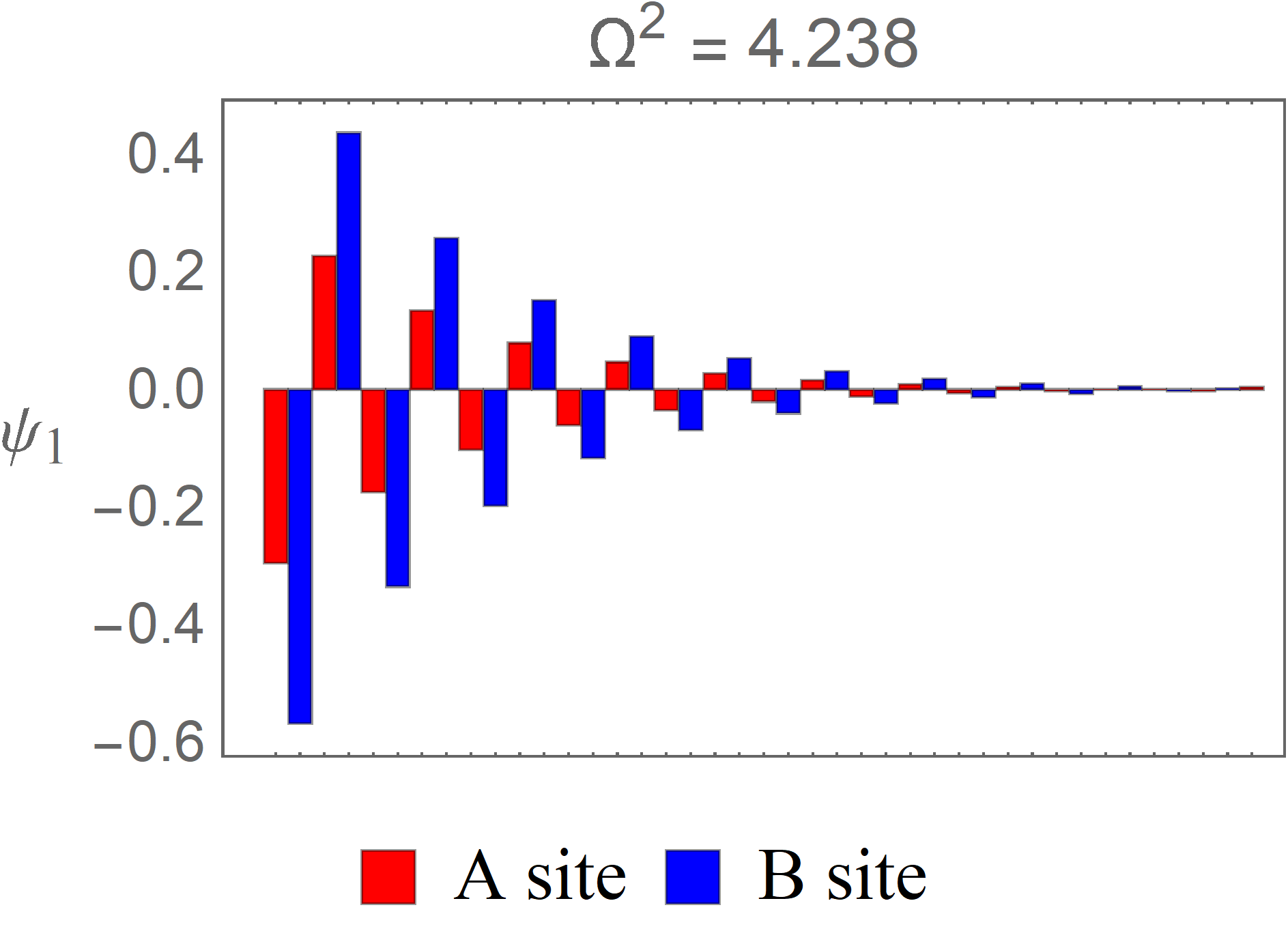}}\\
\subfloat[]{\includegraphics[width=0.40\textwidth]{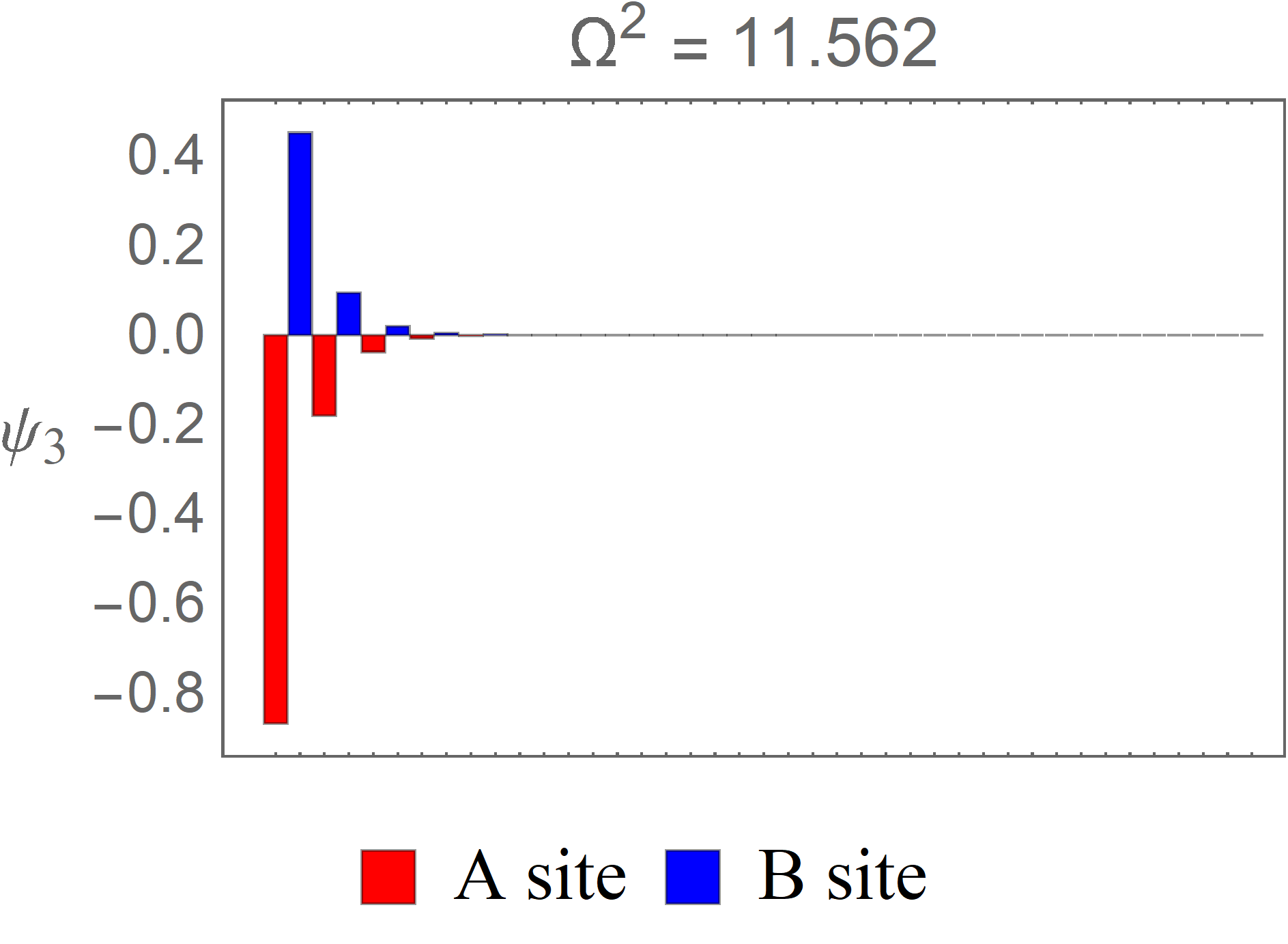}}\hskip 0.5cm
\subfloat[]{\includegraphics[width=0.40\textwidth]{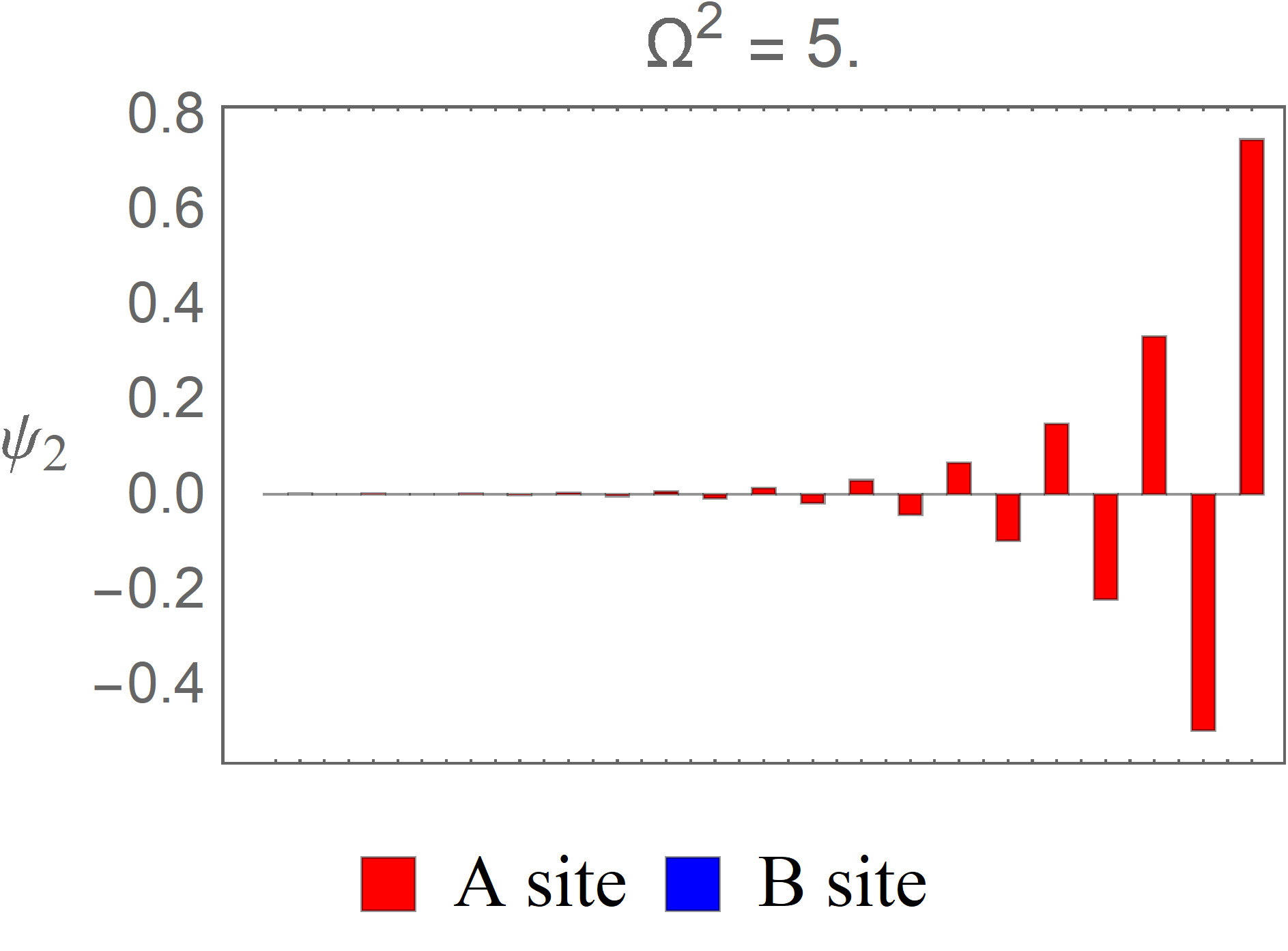}}\\
\subfloat[]{\includegraphics[width=0.40\textwidth]{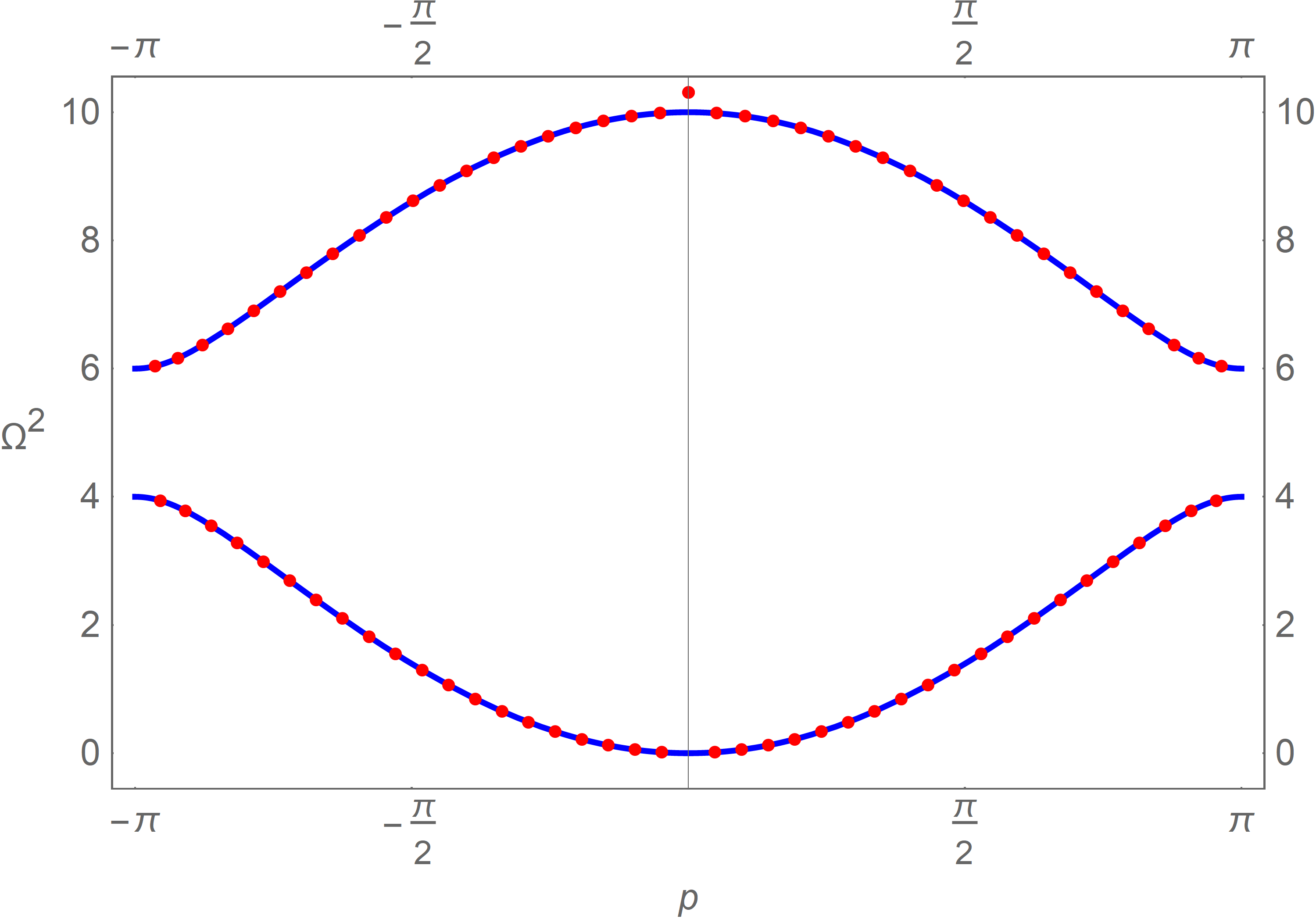}}\hskip 0.5cm
\subfloat[]{\includegraphics[width=0.40\textwidth]{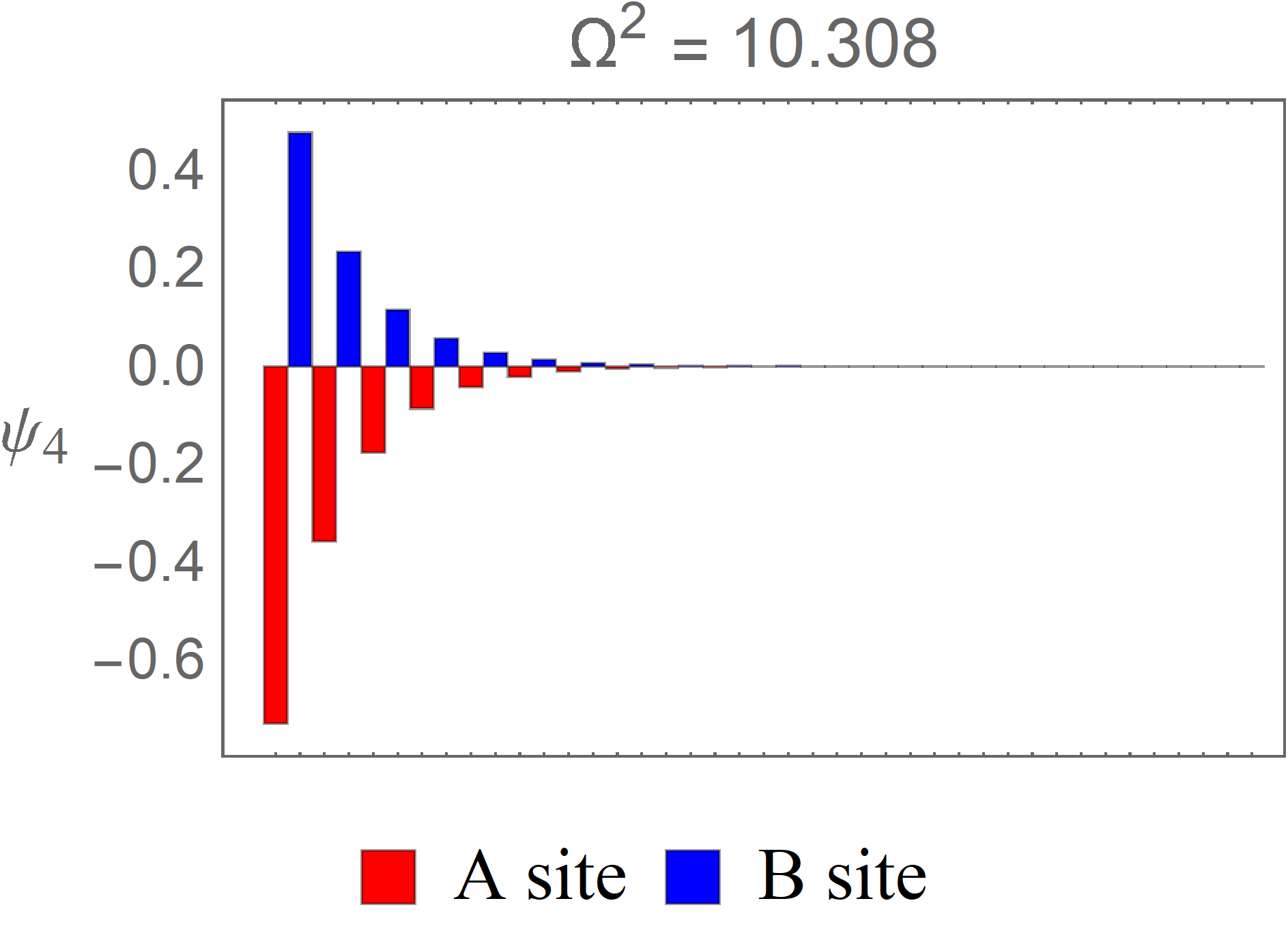}}\\
\caption{The total number of sites $N_{\rm tot} = 41$ and $\Kh=7>2\kh_1$. $\left( \kh_0, \kh_1\right) = (3, 2)$ in (a), (b), (c), and (d), while $\left( \kh_0, \kh_1\right) = (2, 3)$ in (e) and (f). Hence, the system is in the trivial and topological phases, respectively. (a) The energy spectrum of the system for $\left( \kh_0, \kh_1\right) = (3, 2)$.  Note that there are three edge states in total. Two of them are left edge states, with one mid-gap state near the top of the lower band and the other well above the upper band.  In addition, there is also a mid-gap right edge state, which appears because the ``winding number'' for the right boundary is one as $N_{\rm tot}$ is odd. It is not relevant for our study here.  (b), (c) The wave functions of the two left edge states, $\psi_1, \psi_3$, in the system. Note that the ``wave functions'' of the left edge states are non-vanishing both on the A and B sites in contrast to that of the right edge states. (d) The wave functions of the right edge mig-gap state, $\psi_2$. (e) The energy spectrum of the system for $\left( \kh_0, \kh_1\right) = (2, 3)$.  Note that there is only a left edge state with energy above the top of the upper band. (f) The wave functions of the edge state $\psi_4$. }  \label{fig4 SSH-spring}
\end{figure}

\begin{figure}[hbt!]
\centering
\subfloat[]{\includegraphics[width=0.40\textwidth]{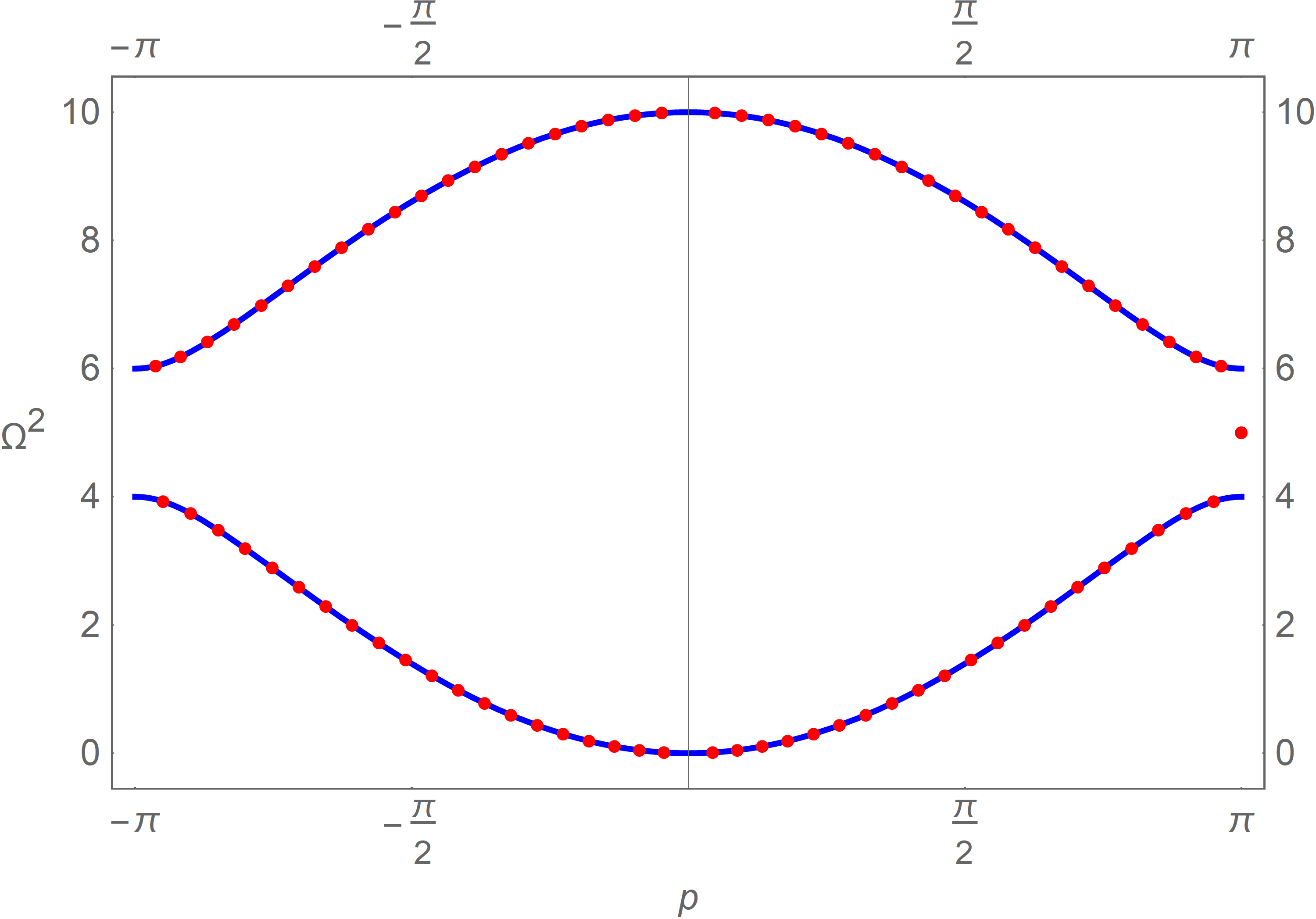}}\hskip 0.5cm
\subfloat[]{\includegraphics[width=0.40\textwidth]{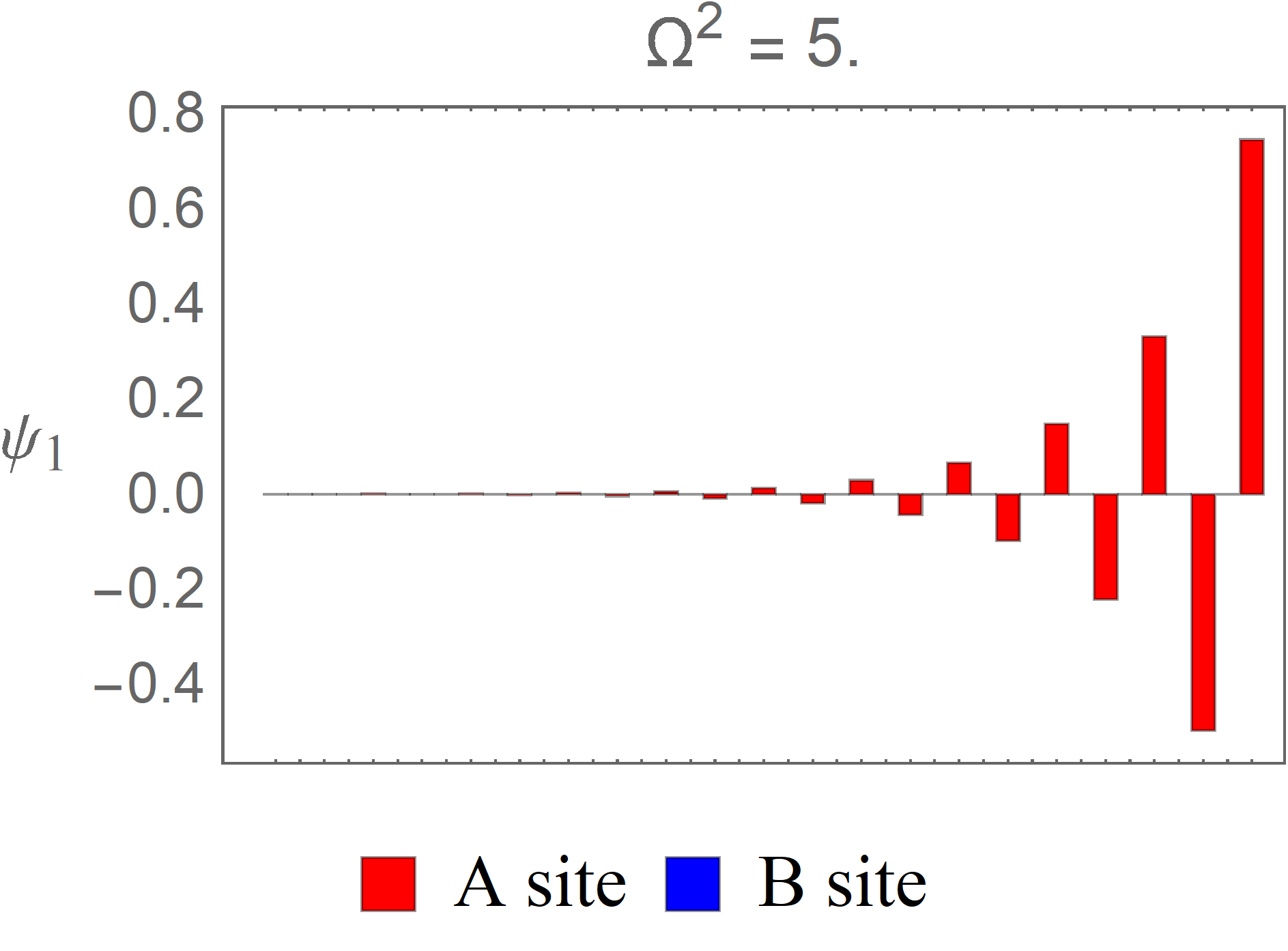}}\\
\subfloat[]{\includegraphics[width=0.40\textwidth]{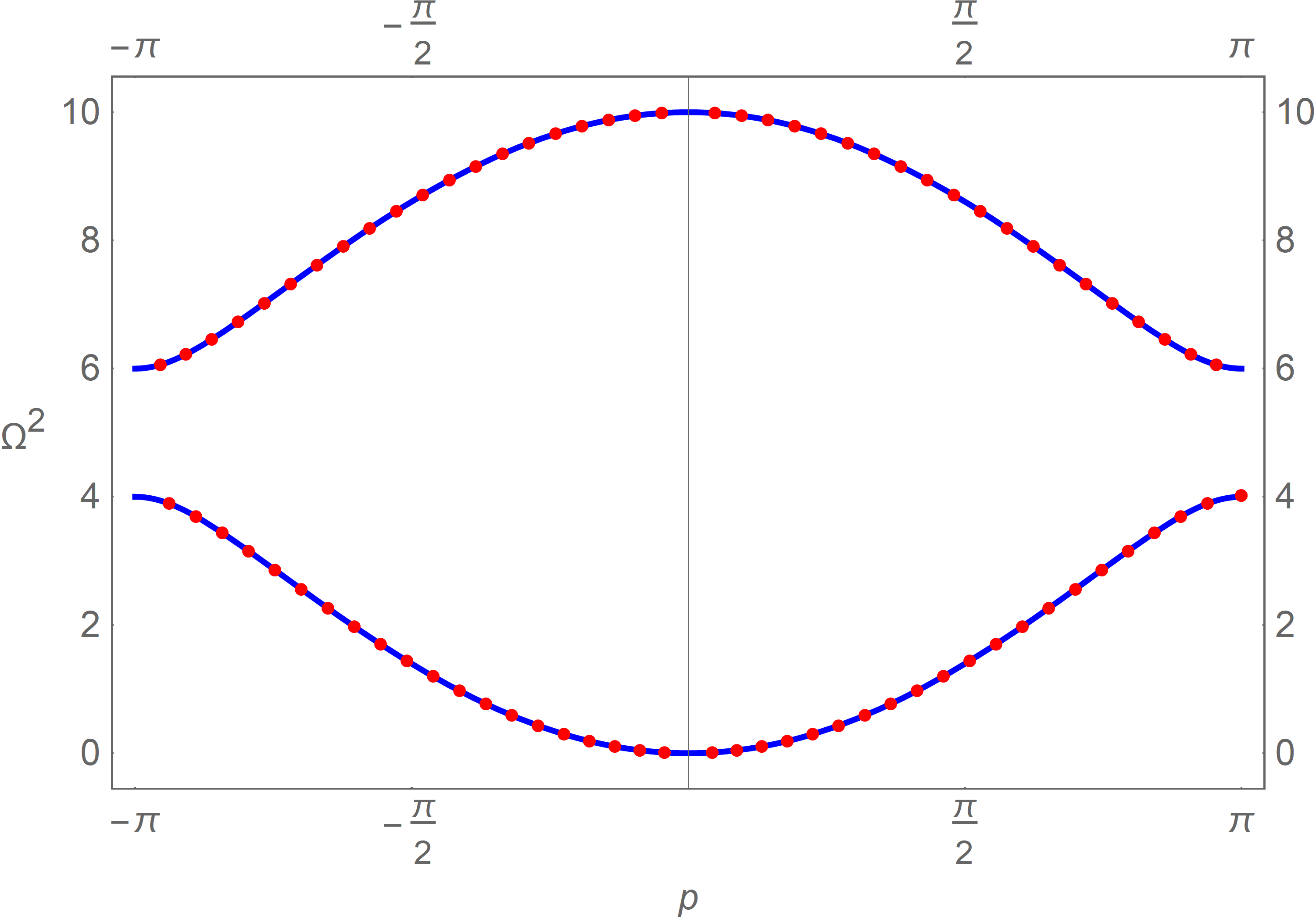}}\hskip 0.5cm
\subfloat[]{\includegraphics[width=0.40\textwidth]{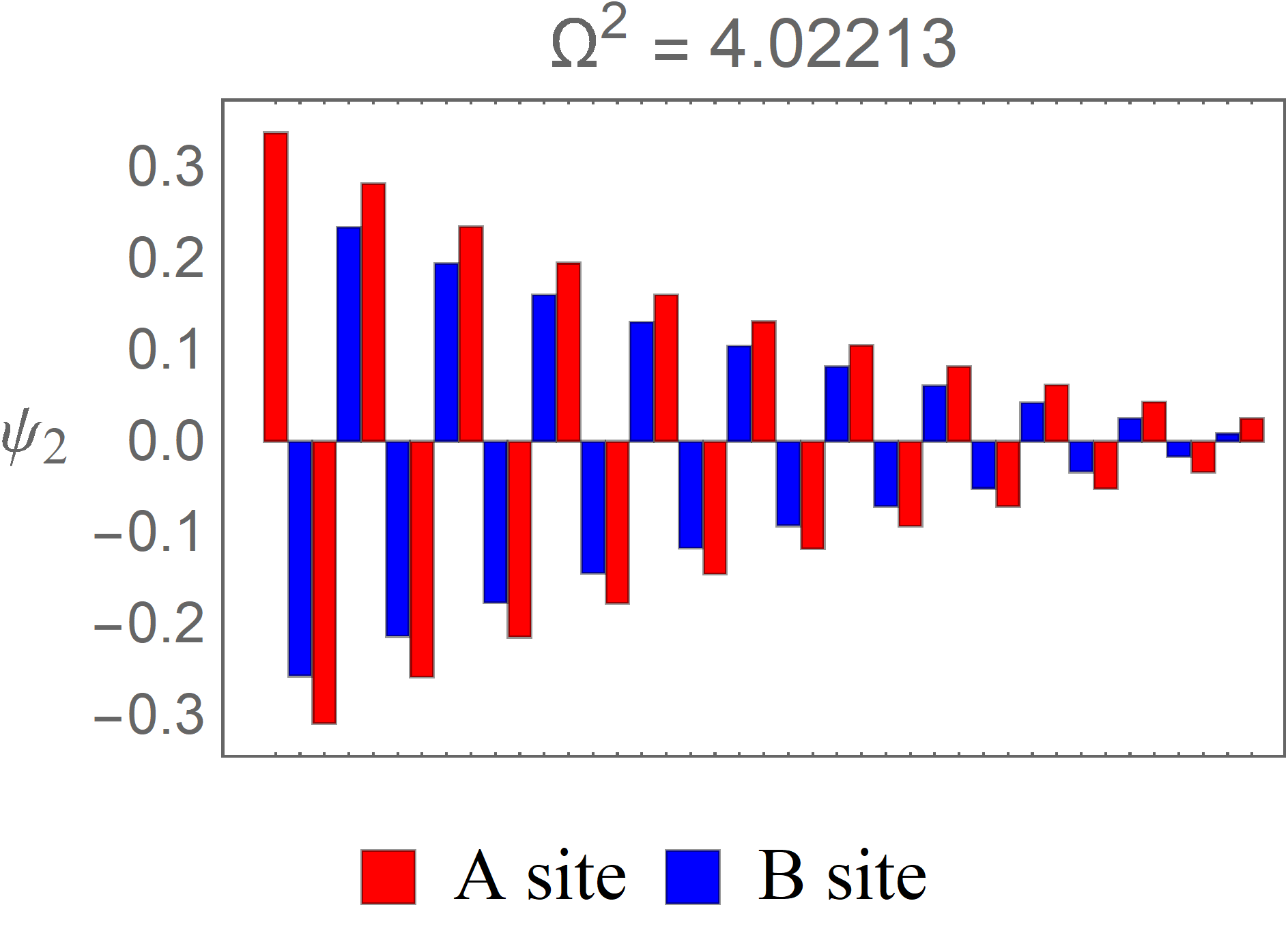}}\\
\caption{The total number of sites $N_{\rm tot} = 41$ and $\Kh=1/2<2\kh_1$. $\left( \kh_0, \kh_1\right) = (3, 2)$ in (a) and (b), while $\left( \kh_0, \kh_1\right) = (2, 3)$ in (c) and (d). Hence, the system is in the trivial and topological phases, respectively. (a) The energy spectrum of the system for $\left( \kh_0, \kh_1\right) = (3, 2)$.  There is only a right edge state.  (b) The wave function of the mid-gap right edge state $\psi_1$ is shown.  It exists for the same reason explained in Fig.~4d. (c) The energy spectrum of the system for $\left( \kh_0, \kh_1\right) = (2, 3)$.  Note that there is only a left edge state with energy near the top of the lower band. (d) The wave functions of the left edge state $\psi_2$. }  \label{fig5 SSH-spring}
\end{figure}

Here, we choose $N_{\rm tot} = 41$ so that there is an odd number of sites.  As a result, the ``winding number'' of the right boundary is related to that of the left boundary by $\n_{\rm R} = 1 - \n_{\rm L}$ so the right boundary will be in the opposite phase of that of the left boundary \cite{Ext-SSH}.  In the trivial and topological phases, we choose $\left( \kh_0, \kh_1 \right) = (3, 2)$ and  $\left( \kh_0, \kh_1 \right) =(2, 3)$, respectively so that it is easier for us to make comparison with the previous results. We show the results for $\Kh =7>2\kh_1$ and $\Kh =1/2<2\kh_1$ in Fig.~\ref{fig4 SSH-spring} and Fig.~\ref{fig5 SSH-spring}, respectively.

\begin{figure}[hbt!]
\centering
\subfloat[]{\includegraphics[width=0.40\textwidth]{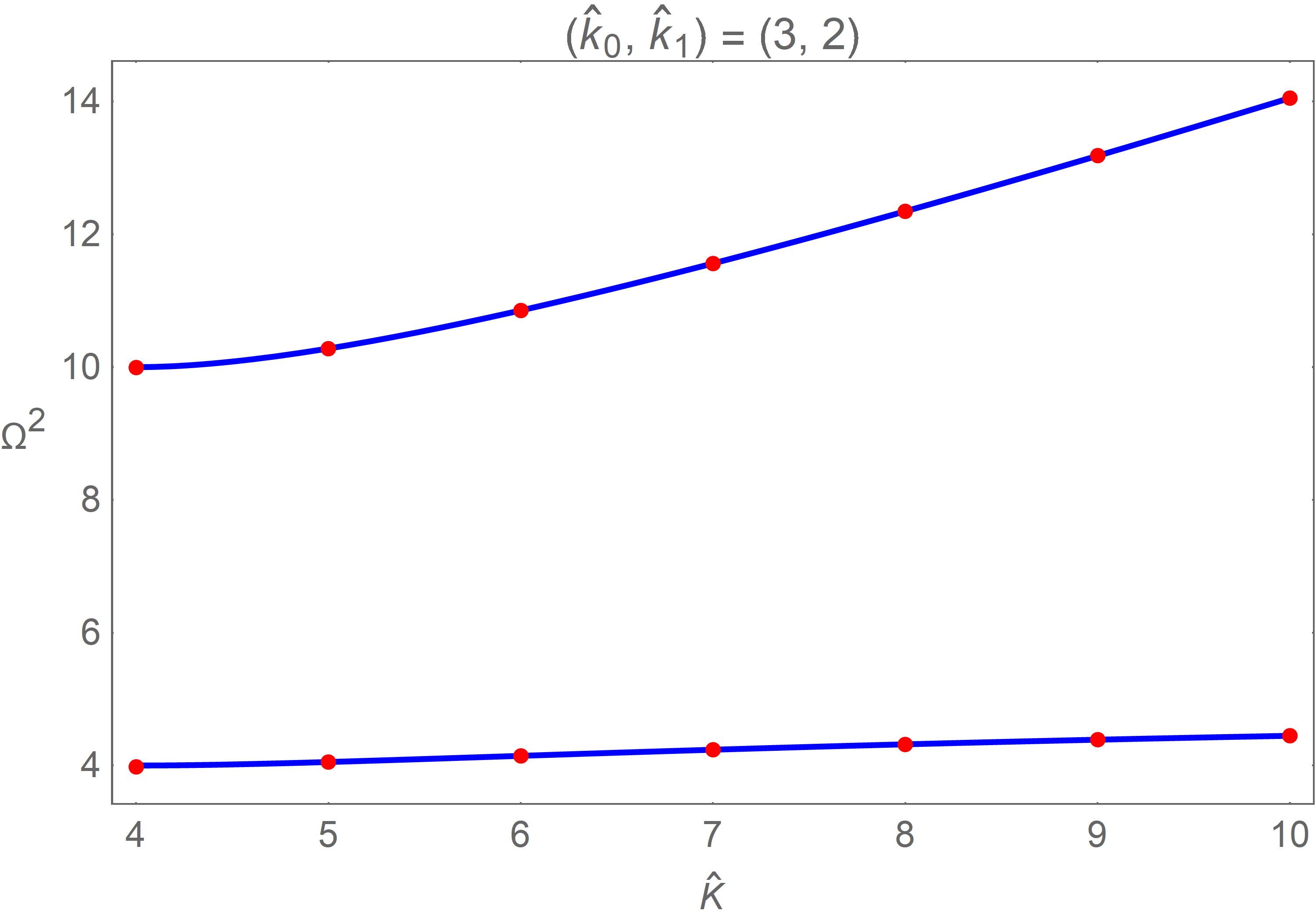}}\hskip 0.5cm
\subfloat[]{\includegraphics[width=0.40\textwidth]{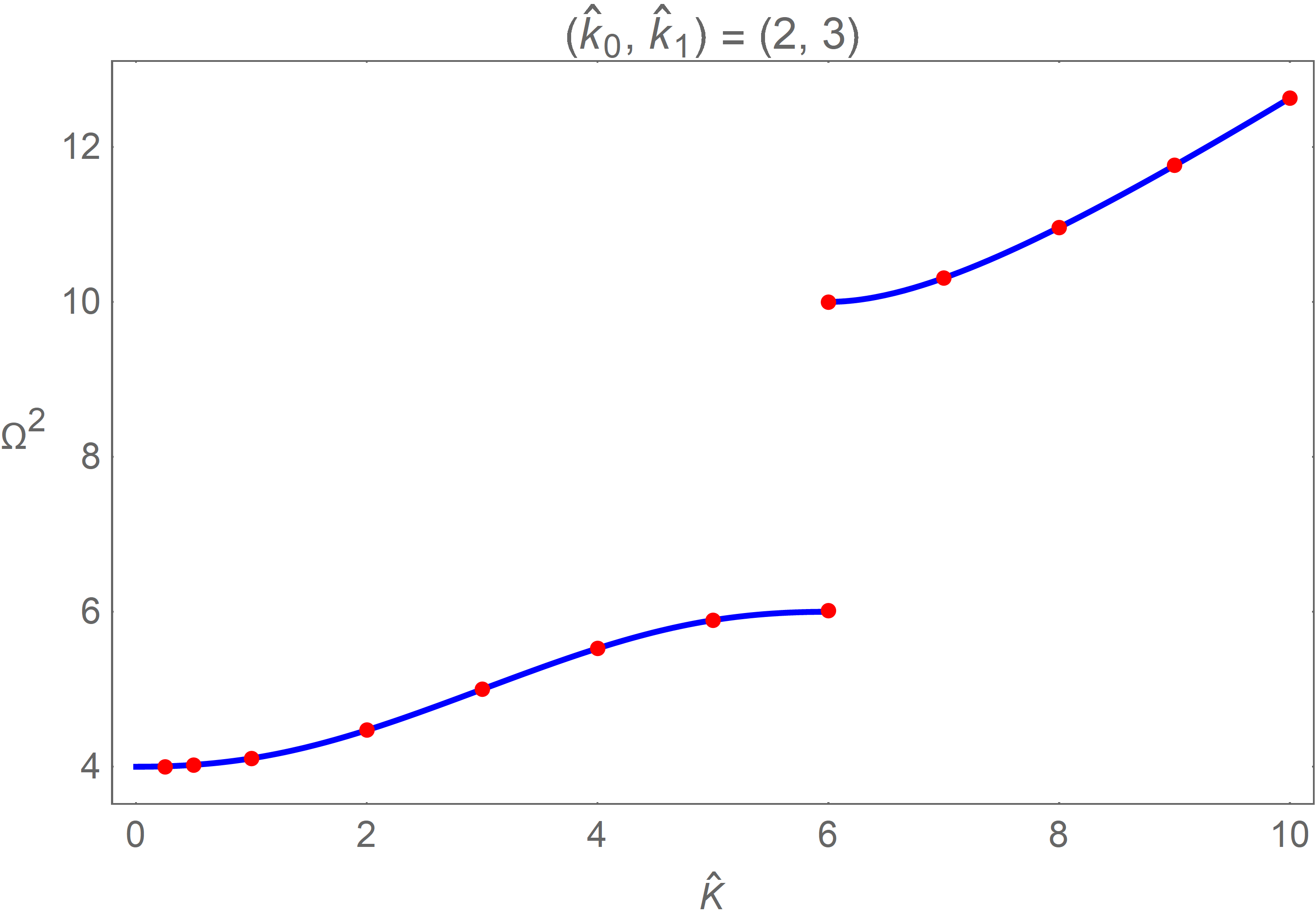}}\\
\caption{$\Om^2$ of the left edge states versus $\Kh$ is plotted. Here, $\Kh$ is proportional to the force constant of the spring attached to $a_1$ from the left. The left boundary of the system satisfies the fixed-end BC and the total number of sites $N_{\rm tot} = 41$. The system is in the trivial and topological phases in (a) and (b), respectively. The results for a right semi-infinite chain are indicated in a solid blue curve and the results of the finite chain are indicated in red dots. (a) $\left( \kh_0, \kh_1\right) = (3, 2)$.  For $\Kh<2\kh_1$, there is no left edge state. Hence, we only show the region $\Kh>2\kh_1$, where there are two left edge states. One of them is a gap state and the other has energy above the top of the upper band. (b) $\left( \kh_0, \kh_1\right) = (2, 3)$.  Note that there is a left edge gap state for $\Kh<2\kh_1$. The energy of the edge state is above the top of the upper band for $\Kh>2\kh_1$.  }  \label{fig6 SSH-spring}
\end{figure}

We also compare the energies of the left edge states in a finite chain to those in a right semi-infinite chain. The result is shown in Fig.~\ref{fig6 SSH-spring}.  One can see that they agree to a high accuracy. In the trivial phase, there are always two edge states. The gap state belongs to the lower band and its energy stays near the top of the lower band.  It is specified by $s= s_-$ in eq.~(\ref{Fixed-end-Edge-s}).  On the other hand, the highest state belongs to the upper band and the corresponding $\Om^2$ grows linearly with $K$ for asymptotic value of $K$. It is given by $s= s_+$ in eq.~(\ref{Fixed-end-Edge-s}).  In the topological phase, the gap state belongs to the lower and upper band for $K\le k_1$ and $k_1< K \le 2k_1$, respectively. The gap state is specified by $s= s_-$.  Again, the highest state belongs to the upper band and the corresponding $\Om^2$ grows linearly with $K$ for large $K$.  It is given by $s= s_+$.

Another possibility is to have a general free-end BC on the left boundary by adding another site, $b_0$, with the force constant of the spring between $b_0$ and $a_1$ being $K$.  The corresponding Hamiltonian for such a right semi-infinite chain is
\bea
&\;& \hskip -1.2cm H^{\rm R}_{{\rm SSH-}K-2}=\sum_{j=1}^{\infty}\Biggl\{ \frac{m}{2} \left[ \dot{a}_j^2(t) + \dot{b}_j^2(t) \right] + \frac{k_0}{2} \left[ a_j(t) - b_j (t) \right]^2 +  \frac{k_1}{2} \left[ a_{j+1}(t) - b_j(t)  \right]^2  \Biggr\} + \frac{m}{2} \dot{b}_0^2(t)  + \frac{K}{2} \left[ a_1(t) - b_0(t) \right]^2.
\eea
When $K \to \infty$, it reduces to the usual free-end BC, $b_0 - a_{1} =0 $. On the other hand, when $K=k_1$, it becomes another free-end BC, $b_0 - a_{-1} =0 $.  At first glance, it seems unlikely that there would be any edge states on the boundary. However, there  do exist edge states to our surprise.

The BC for $b_0$ and $a_1$ are given by
\begin{subequations}
\label{BC SSH-free-R-inf}
\begin{align}
& \Om^2 B_0 + \Kh (A_1 - B_0)  \hskip 2.30cm = 0
\label{BC SSH-free-R-inf-B0} \\
& \Om^2 A_1 + \Kh (B_0 - A_1) + \kh_0 (B_1 - A_1)  = 0.
\label{BC SSH-free-R-inf-A1} \\
& \Om^2 B_1 + \kh_0 (A_1 - B_1) + \kh_1 (A_2 - B_1)  = 0.
\label{BC SSH-free-R-inf-B1}
\end{align}
\end{subequations}
Again, the edge states in this system should be depicted by eq.~(\ref{SSH-s-sol}) with $|s|<1$, with $B_0$ viewed as an independent variable in addition to $\a$ and $\b$.  The secular equation is still given by eq.~(\ref{SSH-secular}). The corresponding characteristic equation of $s$ may be obtained from the condition that the determinant formed by the coefficients of $B_0$, $\a$ and $\b$ is vanishing:
\bea
(\kh_0 + \kh_1 s) \left[\kh_0 \kh_1 + 2 \kh_1^2 - 2 \kh_1 \Kh + \kh_0 (\kh_1 - \Kh) s \right]
- \left[\kh_0 \kh_1 + ( \kh_0 + 2\kh_1)( \kh_1 - \Kh) s \right] \Omh^2 = 0. \nn
\eea
After simplification, it becomes
\bea
f_2(s) = \kh_0 \kh_1^2  - \kh_1 \left(\kh_0 \kh_1 - 2\kh_0\Kh - 4\kh_1\Kh + 4\Kh^2 \right) s - \kh_0 \left(\kh_1 - 3 \Kh \right) \left( \kh_1 - \Kh \right) s^2 + \kh_0 \left( \kh_1 -  \Kh \right)^2  s^3 = 0.
\eea
It is a cubic equation and its roots are given by $s_0, s_1$ and $s_2$, with
\bea
\label{Free-end-Edge-s}
&\;& \hskip -4.5cm   s_J =
-\frac{c_2}{3 c_3}
+ \frac{\left\{ D_2 + \sqrt{4 D_1^3 + D_2^2} \right\}^{1/3}}{3 \sqrt[3]{2} c_3 } e^{i2\pi J/3}
- \frac{\sqrt[3]{2} D_1}{3c_3 \left\{D_2 + \sqrt{4 D_1^3 + D_2^2} \right\}^{1/3} } e^{-i2\pi J/3},
\eea
with $J=0, 1, 2$.  Here, $D_1 = \left(-c_2^2 + 3c_1 c_3 \right)$ and $D_2 =  \left(-2c_2^3 + 9 c_1 c_2  c_3 - 27 c_0 c_3^2 \right)$, with $c_i$ the coefficients of $s^i$ in the polynomial $f_2(s)$, where $i = 0,1, 2, 3$ . The discriminant of the cubic equation is given by:
\bea
\label{discriminant}
&\;& \hskip -4.5cm   4k_0(K - k_1 )^2 \left\{ 4k_0 k_1^2 -12 k_1(k_0 + k_1) K + 3(3 k_0 + 4k_1)K^2 \right\}.
\eea
It is vanishing at $K= K_0$ and $K\pm$, with
\bea
\label{discriminant-root}
&\;& \hskip -4.5cm   K_0 = k_1,\quad  K_\pm  = \frac{ 6k_1(k_0 + k_1) \pm 2 k_1 \sqrt{ 3k_1(2k_0 + 3k_1)} } {3(3k_0 +4 k_1)},
\eea
where double roots of $s$ will show up.  Moreover, we have
\bea
&\;& \hskip -6.5cm
\begin{cases}
f_2(-1)  = &- 4\Kh(\kh_1 - \kh_0) (\kh_1 - \Kh ), \\
f_2(0) \;\;\;  = & \kh_0 \kh_1^2, \\
f_2(1)\;\;\;  = & 2\Kh\left[2\kh_1(\kh_0 + \kh_1) - (\kh_0 + 2 \kh_1) \Kh \right].
\end{cases}
\eea
 From the signs of $f_2(-1), f_2(0)$, and $f_2(1)$ and the information mentioned above, we will be able to tell how many roots have a magnitude less than one. Let's summarize the results here:
\bee[label=\arabic*)]
\item $\kh_0>\kh_1$, the trivial phase:
\bee[label=\roman*)]
\item $\Kh \le  \kh_1$. No edge state.

\item $\kh_1<\Kh< \kh_0$. One edge state with energy between the top of the lower band and the mid-gap energy.

\item $\Kh> \kh_0$. Two edge states. One of them has energy between the top of the lower band and the mid-gap and the bottom of the upper band. The other lies above the top of the upper band.
\eee

\item $\kh_0<\kh_1$, the topological phase:
\bee[label=\roman*)]
\item $\Kh <  \kh_1$. One edge state with energy between the top of the lower band and the bottom of the upper band.

\item $\kh_1 < \Kh < \Kh^*$, with $\Kh^* = 2\kh_1(\kh_0 + \kh_1)/ (\kh_0 + 2 \kh_1)$. No edge state.

\item $\Kh> \Kh^*$. One edge state with energy lying above the upper band.
\eee

\eee
Note that as $K$ decreases, the energy of the gap state approaches the top of the lower band. When $K= 0$, the edge state disappears since the system reduces to the one with a free-end BC in this limit.

Again, one must use a finite chain to check the results numerically. To be specific, we will consider the case that the site by the right boundary satisfies the fixed-end BC.  For $N_{\rm tot} = 2N+1$, the characteristic equation of $s$ for such a system is given by
\bea
&\;& \hskip -1.1cm  \pm  \sqrt{ \kh_0^2 + \kh_1^2 + 2\kh_0 \kh_1 u }
\left\{ \kh_0 \kh_1 \left( \kh_0 + 2\kh_1 - 2\Kh \right) \, U_{N}(u)
+\left( \kh_0^2 \kh_1 + \kh_0 \kh_1^2 + 3\kh_1^3 - \kh_0^2 \Kh -2 \kh_1^2 \Kh \right) \, U_{N-1}(u)  \right. \cr
&\;& \hskip +2.4cm \left. + \kh_0 \kh_1 \left( \kh_1 - \Kh \right) \, U_{N-2}(u)   \right \} \cr
&\;& \hskip -1.1cm
+ \left\{ \kh_0^2 \kh_1^2 U_{N+1}(u)
+ \kh_0 \kh_1 \left( \kh_0^2 + \kh_0 \kh_1 + 3\kh_1^2 - \kh_0 \Kh -2 \kh_1 \Kh \right) U_N(u)  \right. \cr
&\;& \hskip -1.1cm  \left.
+ \left( \kh_0^3 \kh_1 + 3 \kh_0^2 \kh_1^2 + \kh_0 \kh_1^3 +  2\kh_1^4 - \kh_0^3 \Kh  - 2\k_0^2 \kh_1 \Kh - 2\kh1^2 \Kh \right) U_{N-1}(u)  + \kh_0\kh_1 \left(\kh_0 + 2\kh_1 \right) \left(\kh_1 -\Kh \right) U_{N-2} (u) \right\} = 0.
\eea
In contrast, if there are $2N+2$ sites so that $a_{N+1}$ is the site by the right boundary, then the characteristic equation is
\bea
&\;& \hskip -2.1cm  \pm  \sqrt{ \kh_0^2 + \kh_1^2 + 2\kh_0 \kh_1 u }
\left\{ \kh_1\left( \kh_0 + 2\kh_1 - 2\Kh \right) \, U_{N}(u)
+ \kh_0 \left( \kh_1 - \Kh \right) \, U_{N-1}(u)   \right \} \cr
&\;& \hskip -2.1cm
+ \left\{ \kh_0 \kh_1^2 U_{N+1}(u)
+ \kh_1\left( \kh_0^2 +\kh_0 \kh_1 + 2\kh_1^2 -\kh_0 \Kh - 2\kh_1 \Kh \right) U_{N}(u)
+ \kh_0\left(\kh_0 + 2\kh_1 \right) \left(\kh_1 -\Kh \right) U_{N-1} (u) \right\} = 0.
\eea

From our numerical calculation, we find that sometimes the energy of the edge states may be very close to the band edge. Consequently, it is difficult to see whether they are edge states or not just by looking at the energy spectrum of the system. Fortunately, it is well-known that the $s$ associated with a bulk state is uni-modular.  Therefore, by solving the characteristic equation of $s$, we have one more index to identify an edge state.
\begin{figure}[hbt!]
\centering
\subfloat[]{\includegraphics[width=0.40\textwidth]{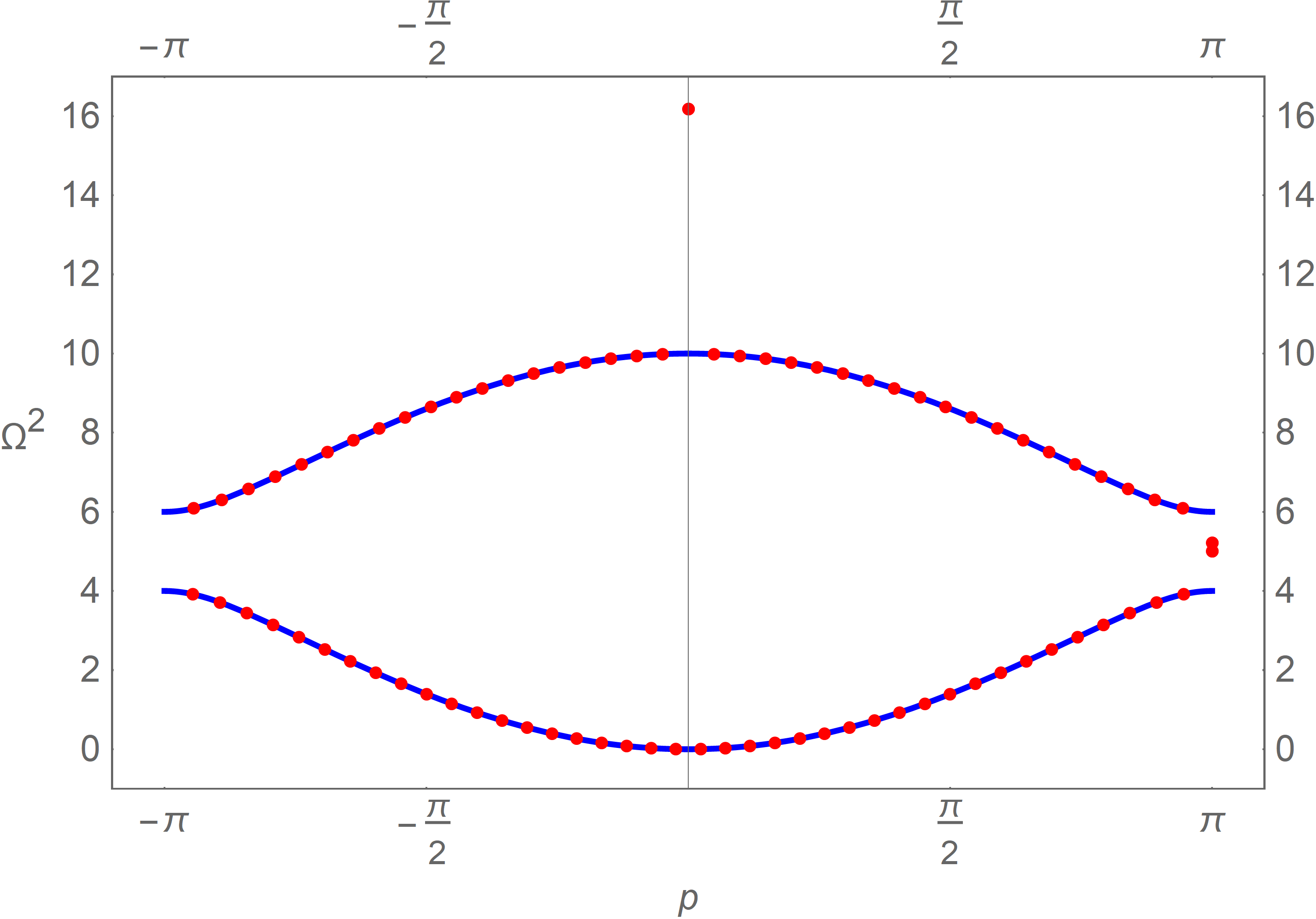}}\hskip 0.5cm
\subfloat[]{\includegraphics[width=0.40\textwidth]{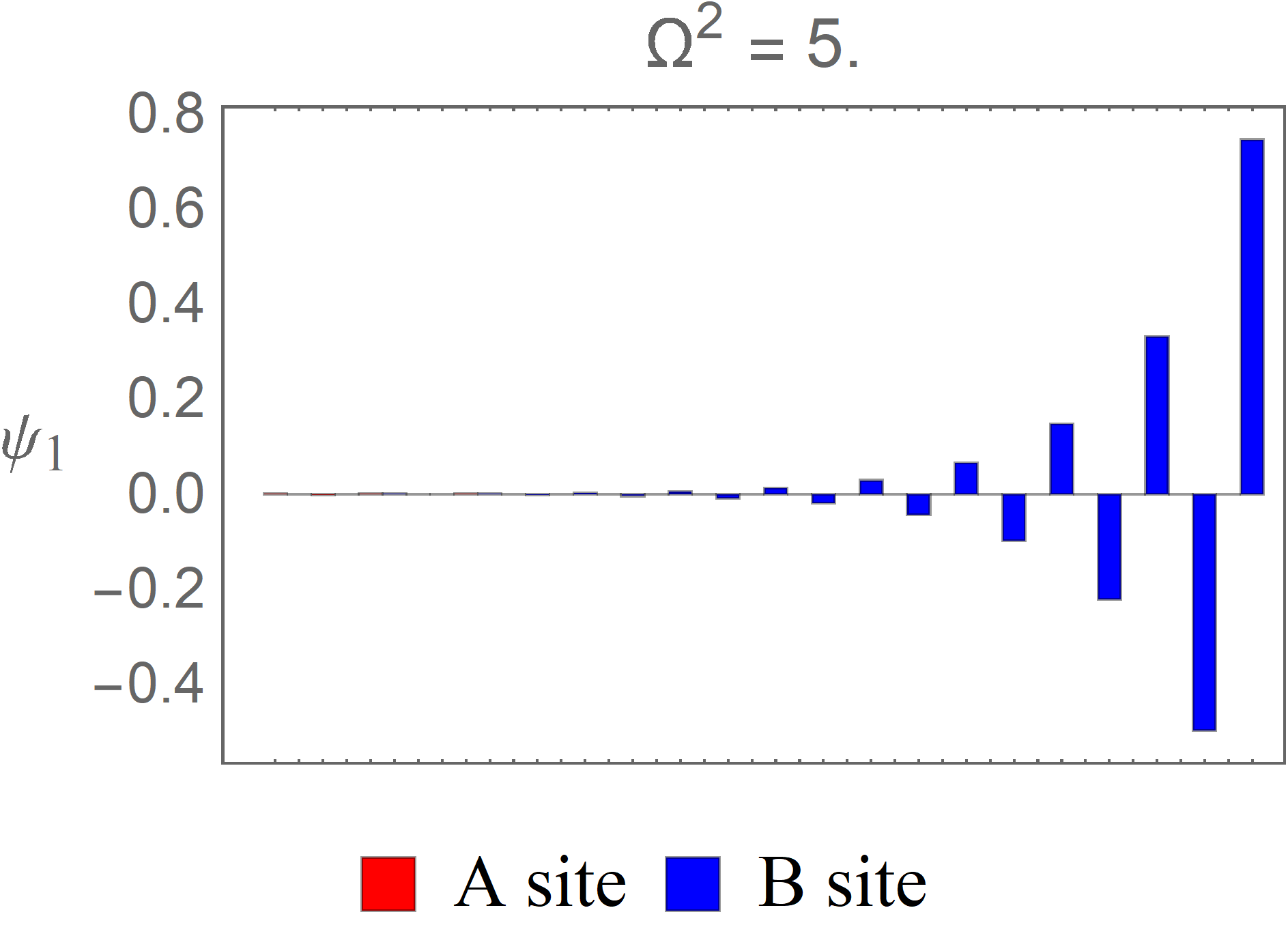}}\\
\subfloat[]{\includegraphics[width=0.40\textwidth]{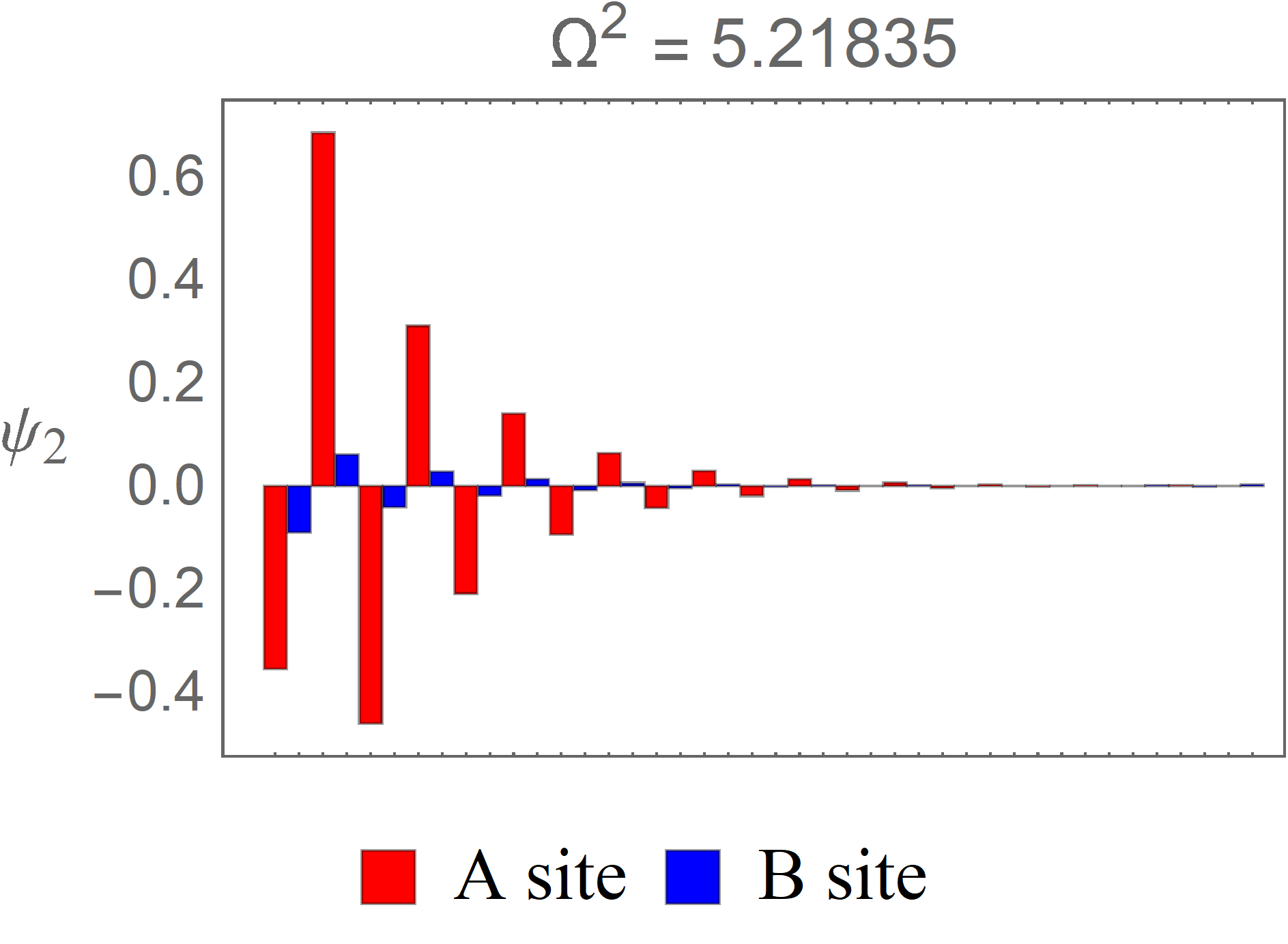}}\hskip 0.5cm
\subfloat[]{\includegraphics[width=0.40\textwidth]{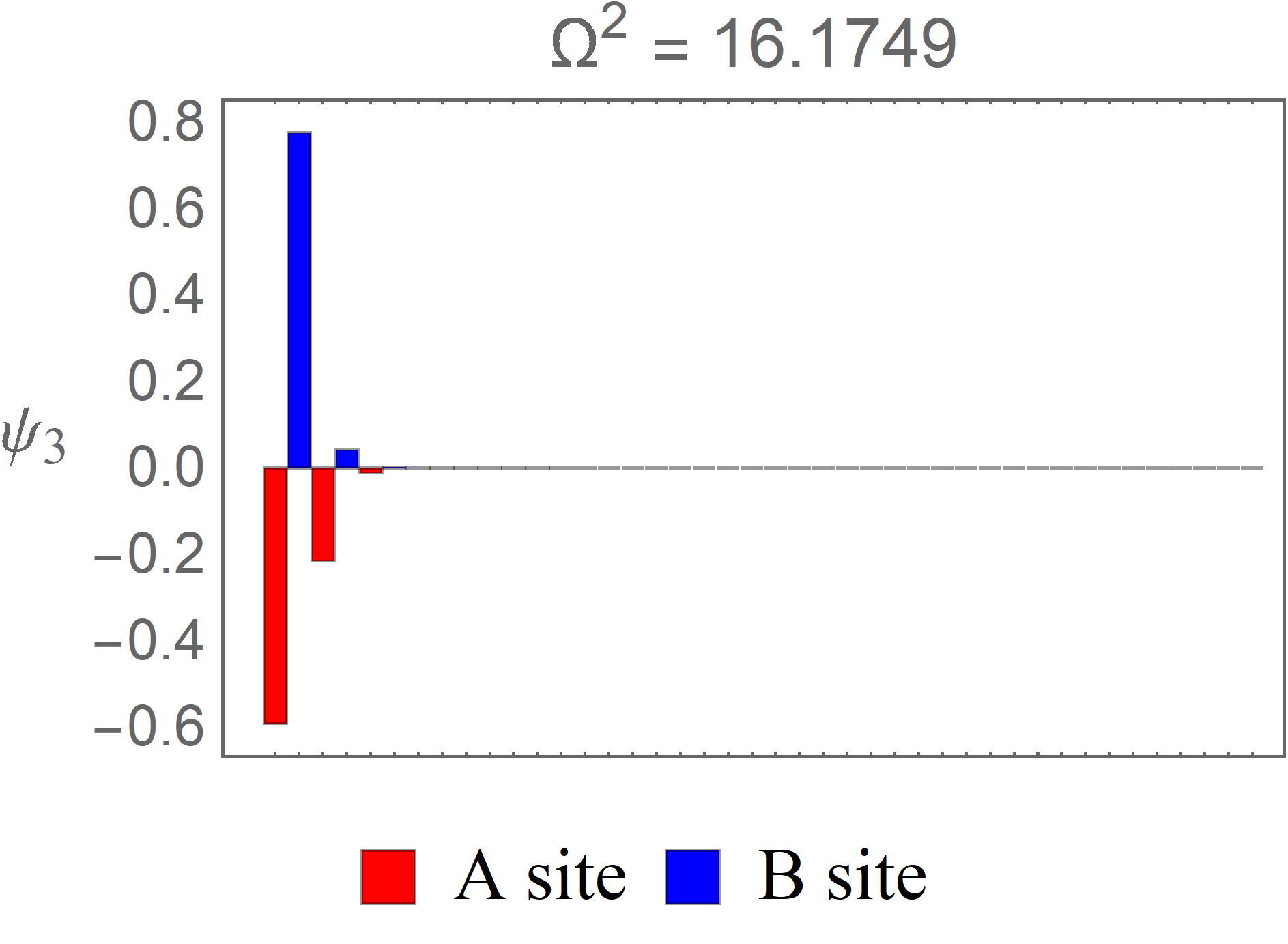}}\\
\subfloat[]{\includegraphics[width=0.40\textwidth]{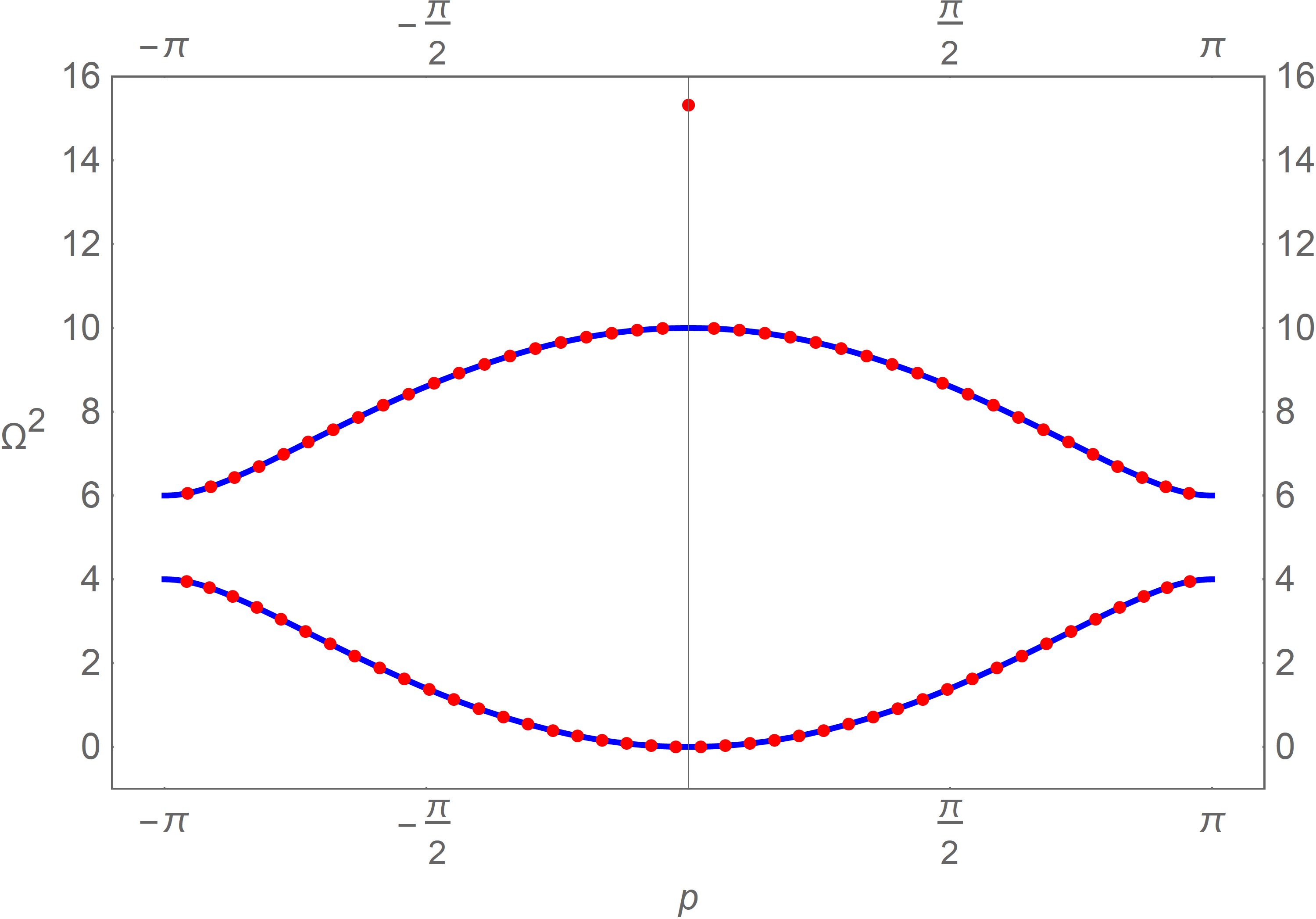}}\hskip 0.5cm
\subfloat[]{\includegraphics[width=0.40\textwidth]{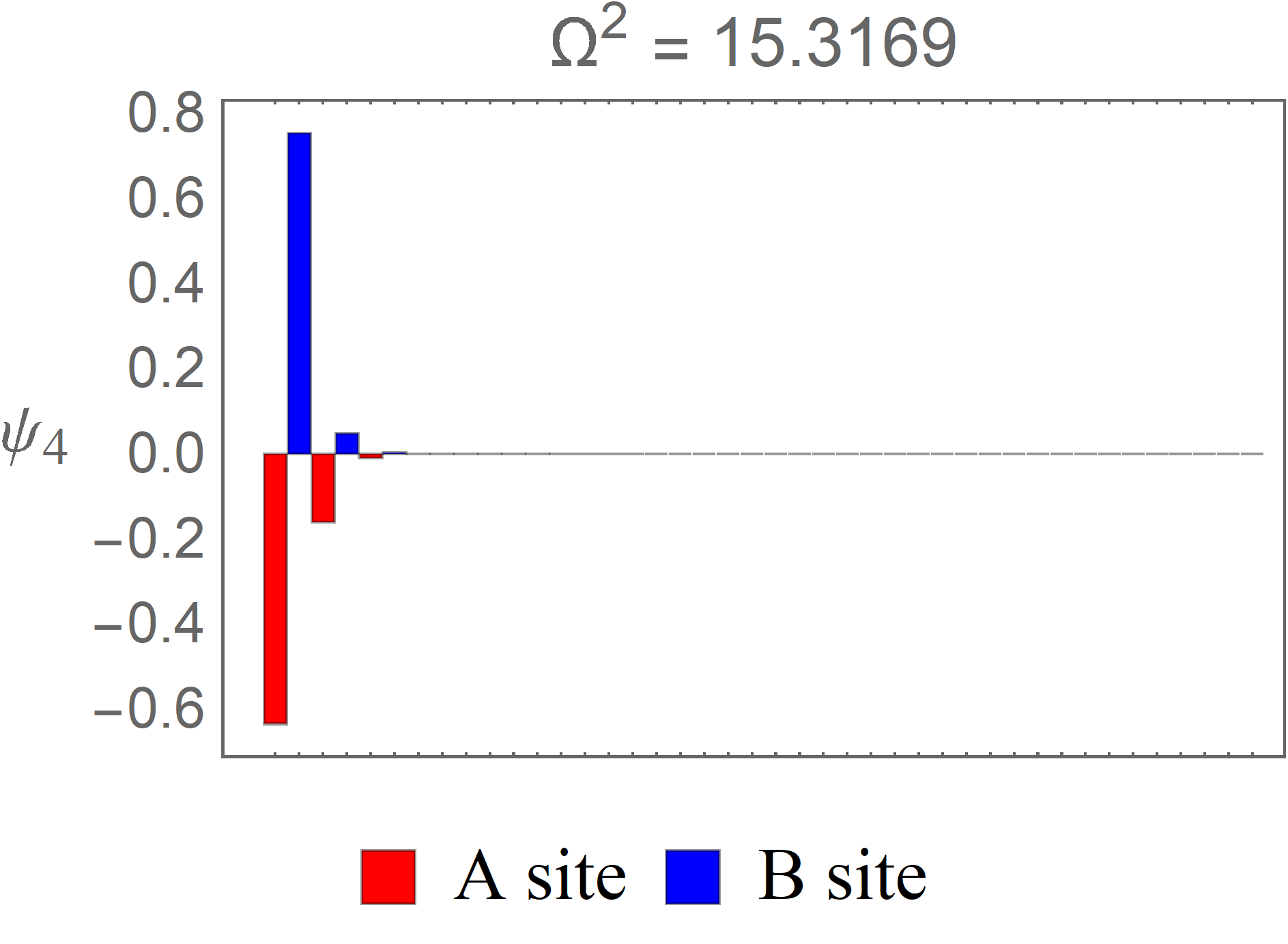}}\\
\caption{The total number of sites $N_{\rm tot} = 42$ and $\Kh=7>2\kh_1$.  $\left( \kh_0, \kh_1\right) = (3, 2)$ in (a), (b), (c), and (d), while $\left( \kh_0, \kh_1\right) = (2, 3)$ in (e) and (f). Hence, the system is in the trivial and topological phases, respectively. (a) The energy spectrum of the system for $\left( \kh_0, \kh_1\right) = (3, 2)$.  Note that there are three edge states in total. Two of them are left edge states, with one gap state near the top of the lower band and the other well above the upper band.  In addition, there is also a mid-gap right edge state, which appears because the ``winding number'' for the right boundary is one as $N_{\rm tot}-1$ is odd. Again, this state is irrelevant for our study here.  (b) The wave functions of the right edge mig-gap state, $\psi_1$. (c), (d) The wave functions of the two left edge states, $\psi_2, \psi_3$, in the system.  (e) The energy spectrum of the system for $\left( \kh_0, \kh_1\right) = (2, 3)$.  Note that there is only a left edge state with energy above the top of the upper band. (f) The wave functions of the edge state $\psi_4$.}  \label{fig7 SSH-spring}
\end{figure}

\begin{figure}[hbt!]
\centering
\subfloat[]{\includegraphics[width=0.40\textwidth]{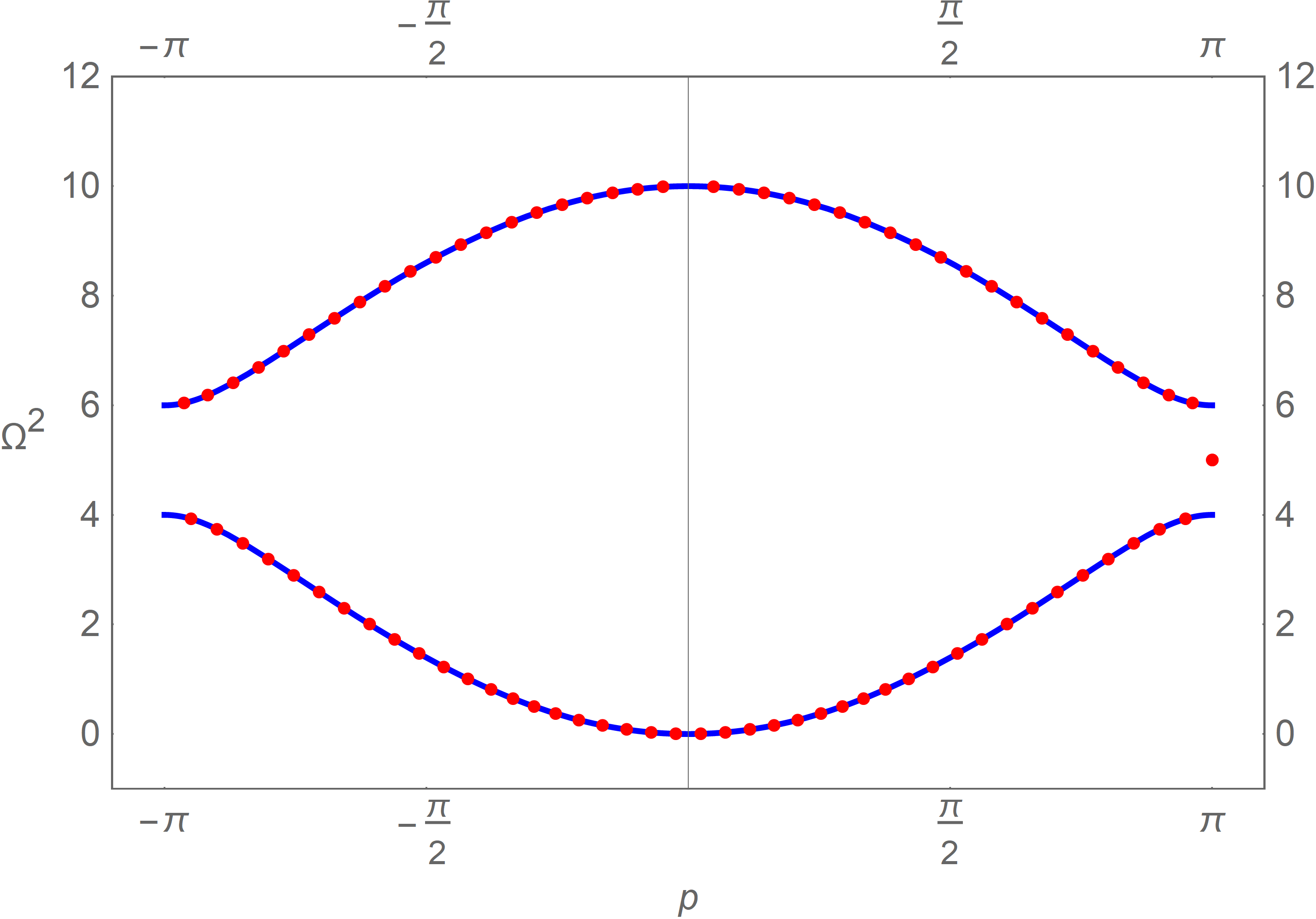}}\hskip 0.5cm
\subfloat[]{\includegraphics[width=0.40\textwidth]{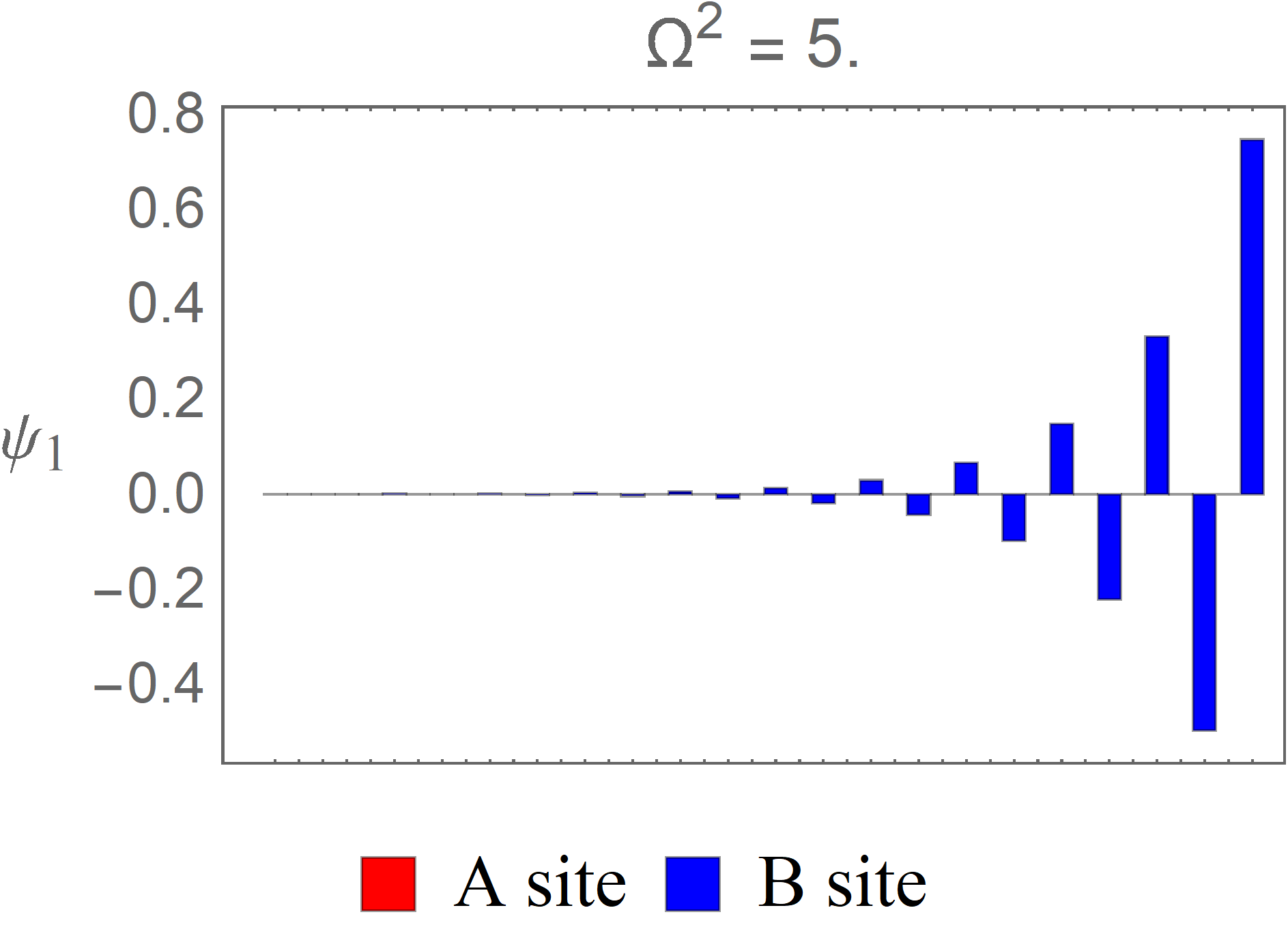}}\\
\subfloat[]{\includegraphics[width=0.40\textwidth]{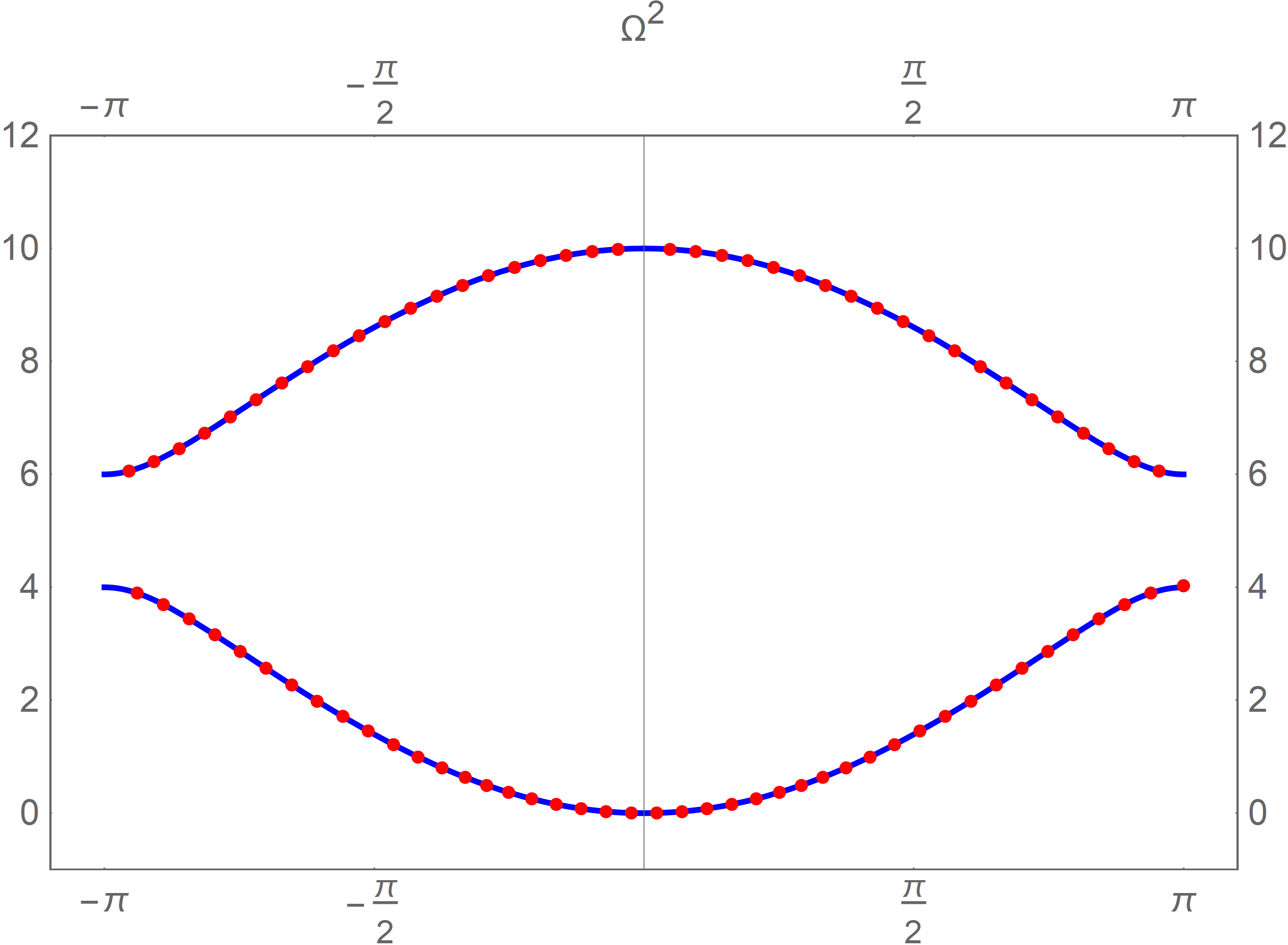}}\hskip 0.5cm
\subfloat[]{\includegraphics[width=0.40\textwidth]{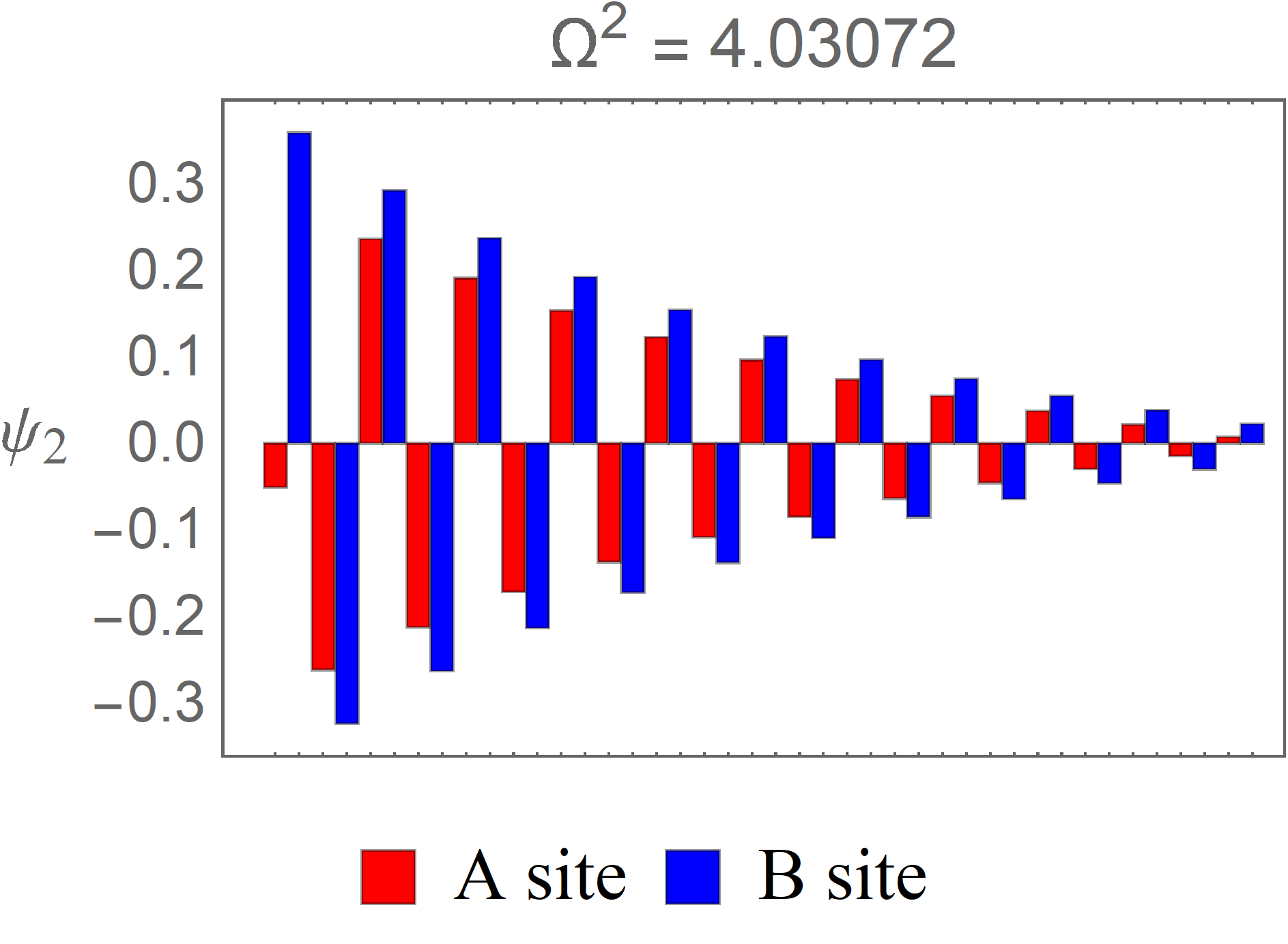}}\\
\caption{The total number of sites $N_{\rm tot} = 42$ and $\Kh =1/2<2\kh_1$.  $\left( \kh_0, \kh_1\right) = (3, 2)$ in (a) and (b), while $\left( \kh_0, \kh_1\right) = (2, 3)$ in (c) and (d). Hence, the system is in the trivial and topological phases, respectively. (a) The energy spectrum of the system for $\left( \kh_0, \kh_1\right) = (3, 2)$.  There is only a right edge state.  (b) The wave function of the mid-gap right edge state $\psi_1$.  (c) The energy spectrum of the system for $\left( \kh_0, \kh_1\right) = (2, 3)$.  Note that there is only a left edge state with energy near the top of the lower band. (d) The wave functions of the left edge state $\psi_2$.}  \label{fig8 SSH-spring}
\end{figure}

\begin{figure}[hbt!]
\centering
\subfloat[]{\includegraphics[width=0.40\textwidth]{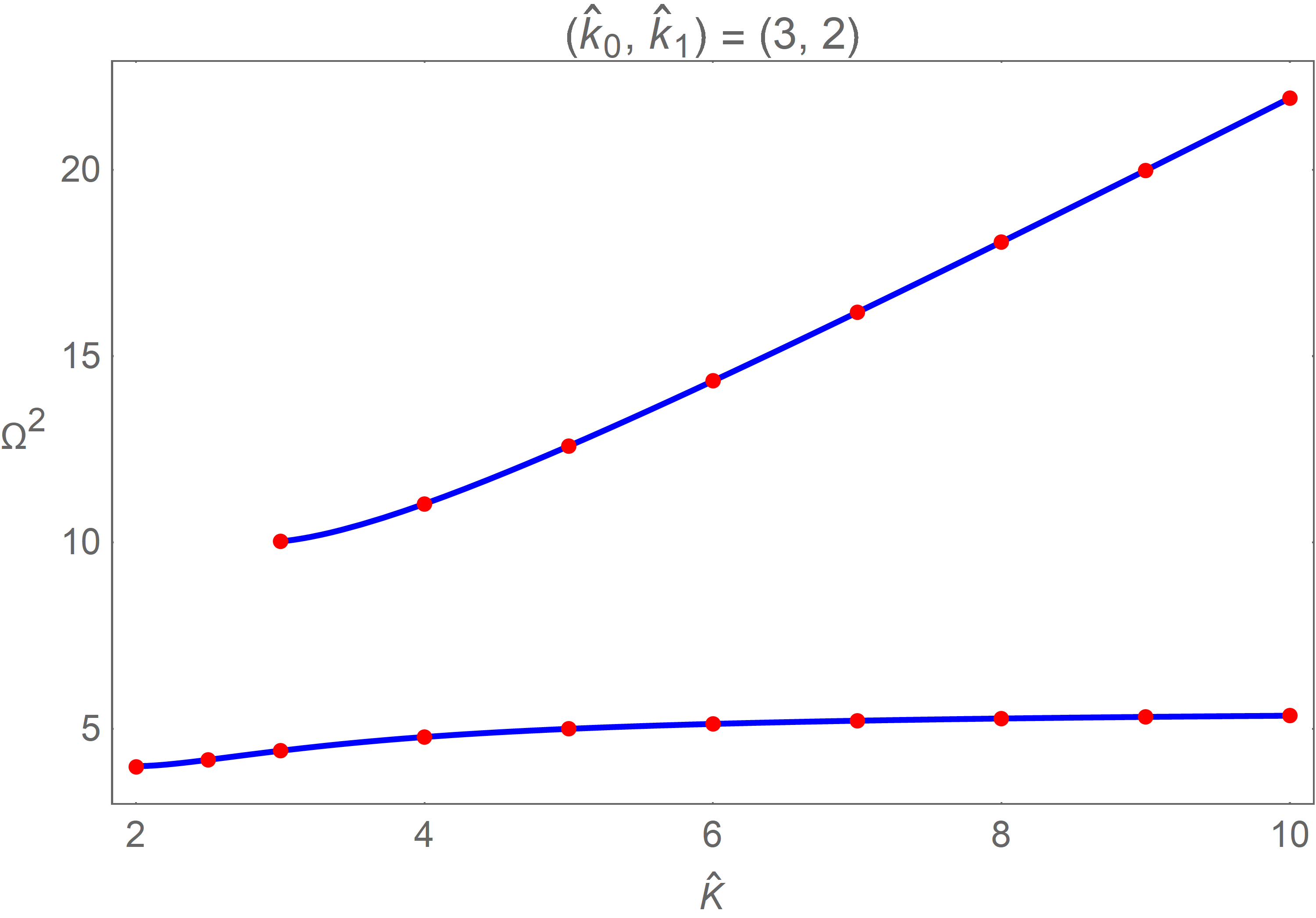}}\hskip 0.5cm
\subfloat[]{\includegraphics[width=0.40\textwidth]{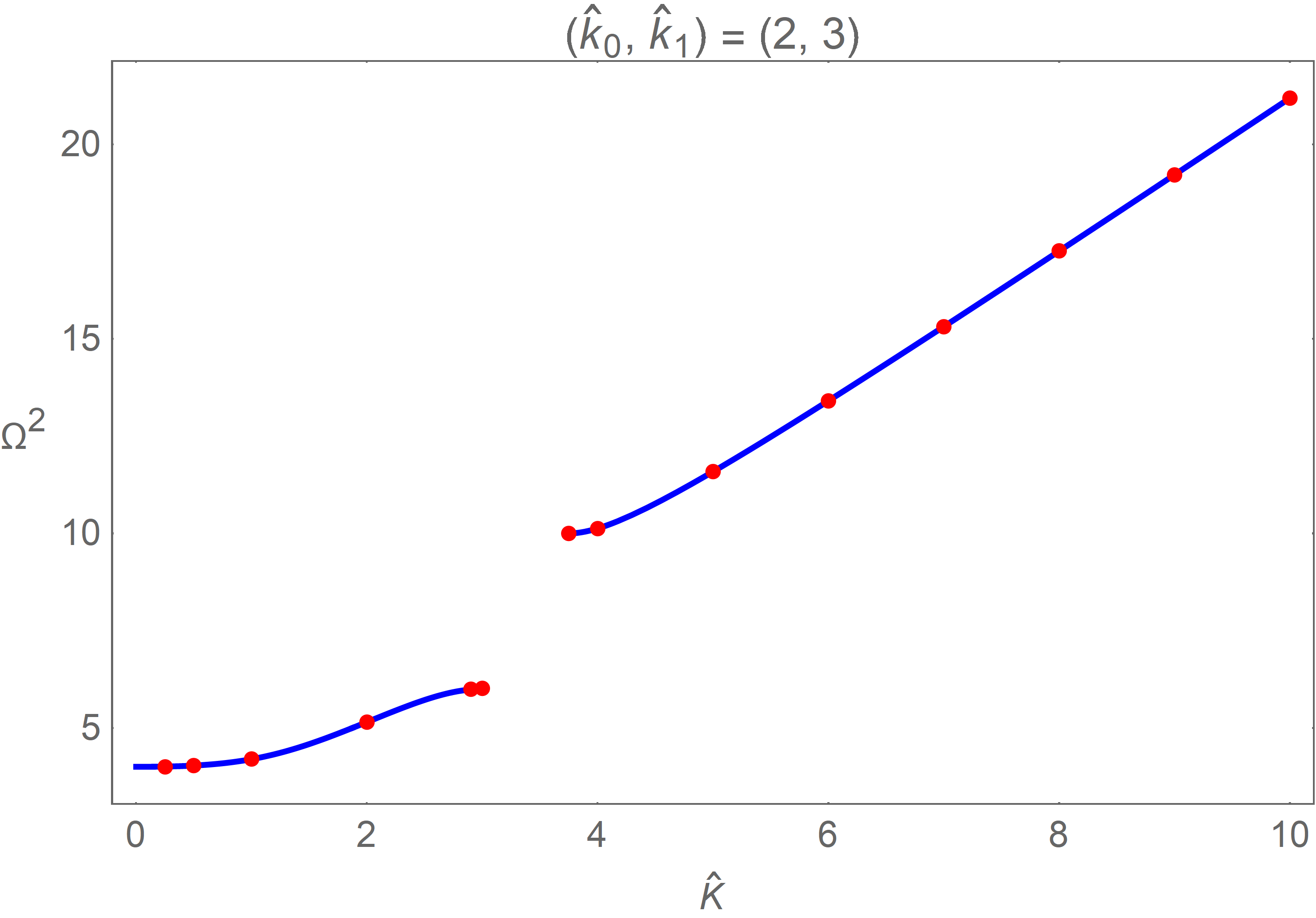}}\\
\caption{$\Om^2$ of the edge states versus $\Kh$ is plotted. Here, $\Kh$ is proportional to the force constant of the spring between $b_0$ and $a_1$. The left boundary of the system satisfies the free-end BC and the total number of sites $N_{\rm tot} = 42$. The system is in the trivial and topological phases in (a) and (b), respectively. The results for a right semi-infinite chain are indicated in a solid blue curve and the results of the finite chain are indicated in red dots. (a) $\left( \kh_0, \kh_1\right) = (3, 2)$.  For $\Kh<\kh_1$, there is no left edge state. Hence, we only show the region $\Kh>\kh_1$. For $\kh_1 < \Kh < \kh_0$, there is only one left edge state with energy near the top of the lower band.  For $\Kh > \kh_0$ there is one more edge state, which has energy above the top of the upper band.  (b) $\left( \kh_0, \kh_1\right) = (2, 3)$.  Note that there is a left edge mid-gap state for $\Kh<\kh_1$ and there is a left edge state with energy above the top of the upper band for $\Kh>2\kh_1(\kh_0 + \kh_1)/ (\kh_0 + 2 \kh_1)$, with the numerical value being 3.75. There is no edge state for $\Kh>2\kh_1(\kh_0 + \kh_1)/ (\kh_0 + 2 \kh_1) $.  }  \label{fig9 SSH-spring}
\end{figure}

We also compare the energies of the left edge states in a finite chain to those in a right semi-infinite chain. The result is shown in Fig.~\ref{fig9 SSH-spring}.  Again, they agree to a high accuracy. In the trivial phase, for $ k_1 < K \le k_0+k_1$ the gap state belongs to the lower band and its energy ranges between the top of the lower band and the mid-gap value $k_0 + k_1$.  For $K > k_0+k_1$, the gap state belongs to the upper band instead. The state is specified by $s = s_1$ for all values of $K$.  On the other hand, the highest state starts to appear at $K=k_0$ and always belongs to the upper band. The corresponding $\Om^2$ grows linearly with $K$ for large $K$ and the state is specified by $s= s_2$.  In the topological phase, the gap state belongs to the lower band and is specified by $s_1$ and $s_0$ for $K<K_-$ and $K_- \le K <  K^*/2$, respectively. It then belongs to the upper band for $K^*/2  \le K\le k_1$ but still is specified by $s= s_0$. In the window $k_1 \le K < K^*$, there is no allowed edge state. For $K\ge K^*$, the edge state becomes the highest state and it is in the upper band. The corresponding $\Om^2$ grows linearly with $K$ for large $K$, and it is specified by $s=s_2$.

\subsection{IV. Conclusion and discussion}

In this paper, we consider the SSH coupled-spring system and its extensions. It has been shown in the literature that there would exist edge modes on the boundaries of such a system if it is in the topological phase, $k_1>k_0$, and satisfies the fixed-end BCs.  In contrast, if the system satisfies the free-end BCs, there is no edge mode, even if it is the topological phase. We confirm the results both analytically and numerically. Two kinds of  generalization are made. We first show by introducing additional third-nearest-neighbor spring coupling, more topological phases would appear similar to the extended SSH models. We then investigate the effect of changing the force constant of the spring by the boundary in the SSH coupled-spring system. Both fixed-end and free-end BCs are considered. In particular, a right semi-infinite chain is used to simplify the analysis so that analytic results may be obtained. They are verified with numerical calculations using a finite chain. The frequencies of the edge modes are shown to agree to a high accuracy. The most surprising result is that when the force constant of the spring by the boundary is modified, then edge modes could exist even if the system satisfies the free-end BCs.  Moreover, edge modes would generally appear independent of whether the bulk of the system is in the topological or trivial phases. In Ref.~\cite{ChenVit2014}, experiments have been carried out to confirm the predictions of the SSH coupled-spring system. It should not be difficult to modify the setup to check the results we achieved.

Now, we would like to discuss some possible directions for further study. Since the equations that appear in the SSH coupled-spring system satisfying the fixed-end BCs are equivalent to those in the SSH model satisfying the open BCs, the results we obtain here may be carried over to the SSH model. Thus, similar edge states should also exist in the SSH model if we modify the hopping amplitudes between the boundary site and the site next to it.  Since the SSH  has been realized in laboratories using various different setups, we believe our predictions may also be verified in similar setups experimentally.

Just like the SSH model, it is straightforward to generalize the SSH coupled-spring system so that there are more than two bands. This may be achieved both theoretically and experimentally. It is known that in a multi-band SSH model, there may exist gap states that are not protected by the chiral symmetry. Hence, we are not sure whether these states are robust~\cite{Multi-band3}. By modifying the setups used in ref.~\cite{ChenVit2014}, we should be able to settle the issue. Mechanical analogs of topological materials in higher dimensions remain an intriguing topic, especially topological semimetals with Dirac or Weyl points.

\section*{Acknowledgments}
The work is supported in part by the Grants MOST-110-2112-M-A49 -012 -MY2 of the Ministry of Science and Technology, Taiwan.  The authors would like to thank Prof. Ming-Che Chang and Jhih-Shih You for stimulating discussions.

\appendix
\section{Appendix A: Extended SSH coupled-spring system}\label{Ext-SSH}

Here, we consider the extended SSH coupled-spring system, in which there is also the third nearest neighbor coupling.  There are two types of extended SSH coupled-spring systems and their Hamiltonians are given by
\bea
&\;& \hskip -1.5cm H_{\rm SSH-ext1}=\sum_{j=\infty}^{\infty}\Biggl\{ \frac{m}{2} \left[ \dot{a}_j^2(t) + \dot{b}_j^2(t) \right] + \frac{k_0}{2} \left[ a_j(t) - b_j (t) \right]^2 +  \frac{k_1}{2} \left[ a_{j+1}(t) - b_j(t)  \right]^2  +  \frac{k_2}{2} \left[ a_{j+2}(t) - b_j(t)  \right]^2 \Biggr\},
\eea
and
\bea
&\;& \hskip -1.5cm H_{\rm SSH-ext2}=\sum_{j= -\infty}^{\infty}\Biggl\{ \frac{m}{2} \left[ \dot{a}_j^2(t) + \dot{b}_j^2(t) \right] + \frac{k_0}{2} \left[ a_j(t) - b_j (t) \right]^2 +  \frac{k_1}{2} \left[ a_{j+1}(t) - b_j(t)  \right]^2  +  \frac{k_{-1}}{2} \left[ a_{j-1}(t) - b_j(t)  \right]^2 \Biggr\},
\eea
respectively. These two types of systems were first introduced in the context of SSH-like systems.  Again, we will borrow the winding numbers $\n$ to classify them. In these systems, the winding numbers may take the values $(2, 1, 0)$ and $(1, 0, -1)$, respectively.  It is known in the literature that there exists a one-to-one mapping between them. In particular, the former Hamiltonian may be mapped to the latter one by renaming $a_j(t), b_j(t)$ as $\bt_j(t), \at_{j+1}(t)$, and identifying $(k_0, k_1, k_2)$ as $(\kt_1, \kt_0, \kt_{-1})$ \cite{Ext-SSH}.  This may be further confirmed by checking the condition for various phases.  For type 1,
\bea
\label{Type 1 winding number}
&\;& \hskip -2.5cm \n=
\begin{cases}
2, & \mbox{for } k_0 + k_2 > k_1, \mbox{ and }  k_2>k_0 \\
1, & \mbox{for } k_0 + k_2 < k_1; \\
0, & \mbox{for } k_0 + k_2 > k_1, \mbox{ and }  k_2<k_0.
\end{cases}
\eea
After the mapping, we have
\bea
\label{Type 1 winding number}
&\;& \hskip -2.5cm \n=
\begin{cases}
-1, & \mbox{for } k_1 + k_{-1} > k_0, \mbox{ and } k_{-1}> k_1 \\
\;\;\; 0, & \mbox{for } k_1 + k_{-1}< k_0; \\
\;\;\; 1, & \mbox{for } k_1 + k_{-1} > k_0, \mbox{ and } k_{-1}<k_1.
\end{cases}
\eea
Here, we have suppressed the tilde to simplify the symbols.

Similar to the SSH coupled-spring system, let's consider the case of fixed-end BC. For a finite chain of type 1 extended SSH coupled-spring system, the Hamiltonian is given by
\bea
&\;& \hskip -0.0cm H_{\rm SSH-ext1}=\sum_{j=1}^{N-2}\Biggl\{ \frac{m}{2} \left[ \dot{a}_j^2(t) + \dot{b}_j^2(t) \right] + \frac{k_0}{2} \left[ a_j(t) - b_j (t) \right]^2 +  \frac{k_1}{2} \left[ a_{j+1}(t) - b_j(t)  \right]^2  +  \frac{k_2}{2} \left[ a_{j+2}(t) - b_j(t)  \right]^2 \Biggr\} \cr
&\;& \hskip 2.7cm  + \frac{m}{2} \left[  \dot{a}_{N-1}^2(t) + \dot{b}_{N-1}^2(t) \right] + \frac{k_0}{2} \left[ a_{N-1}(t) - b_{N-1} (t) \right]^2 + \frac{k_1}{2} \left[ a_{N}(t) - b_{N-1} (t) \right]^2 \\
&\;& \hskip 2.7cm  + \frac{m}{2} \left[  \dot{a}_N^2(t) + \dot{b}_N^2(t) \right] + \frac{k_0}{2} \left[ a_N(t) - b_N (t) \right]^2 + \frac{\left( k_1+k_2 \right) }{2} \left[ a_1^2(t) +  b_N^2(t) \right] + \frac{k_2 }{2} \left[ a_2^2(t) +  b_{N-1}^2(t) \right]. \nn
\eea
The EOM and BC are
\bea \label{SSH_ext1-R-semi-inf}
&\;& \hskip -3.1cm \Omh^2 A_j + \left(\kh_0 B_j + \kh_1 B_{j-1} + \kh_2 B_{j-2} \right) =0; \cr
&\;& \hskip -3.1cm \Omh^2 B_j + \left(\kh_0 A_j + \kh_1 A_{j+1} + \kh_2 A_{j+2} \right) =0,
\eea
and
\bea
&\;& \hskip -3.1cm B_{-1} =B_0 =0,  \quad A_{N+1}= A_{N+2}=0,
\eea
respectively.  Here, $\Omh^2 = \Om^2 - \left(\kh_0 + \kh_1 + \kh_2 \right).$
After simplification, the secular equation and characteristic equation of $s$ are given by
\bea
&\;& \hskip -2.1cm \Omh^2 = \pm \sqrt{\kh_0^2 + \kh_1^2 + \kh_2^2 + 2\kh_0 \kh_2 \left(1 + 2 u_1 u_2 \right) },
\eea
and
\bea  \label{Simplified characteristic eq 1A even}
&\;& \hskip -2.6cm  \left\{ k_0 U_{N+2}(u_1)  + k_1 U_{N+1}(u_1) + k_2  U_{N}(u_1) \right\}
\left\{ k_0 U_{N}(u_2)  + k_1  U_{N-1}(u_2) + k_2  U_{N-2}(u_2) \right\} \cr
&\;& \hskip -2.95cm  +\left\{ k_0 U_{N+2}(u_2)  + k_1 U_{N+1}(u_2) + k_2  U_{N}(u_2) \right\}
\left\{ k_0 U_{N}(u_1)  + k_1  U_{N-1}(u_1) +k_2  U_{N-2}(u_1) \right\} \cr
&\;& \hskip -3.1cm  -2  \left\{ k_{0} U_{N+1}(u_1)  + k_1 U_{N}(u_1) + k_2 U_{N-1}(u_1) \right\}
\left\{ k_{0} U_{N+1}(u_2)  + k_1 U_{N}(u_2) + k_2 U_{N-1}(u_2) \right\} \cr
&\;& \hskip -3.1cm  + 2 \left\{ k_{0}^2 + k_1^2 + k_2^2 - 2k_0 k_2 \left(1 + 2 u_1 u_2 \right) \right\} = 0,
\eea
respectively.  Here,  $u_I= \left( s_I + s_I^{-1} \right)/2$, $I=1,2$ and they are related by
\bea
u_1 + u_2 = - k_1\left(k_0 + k_2\right)/(2 k_0 k_2 ). \nn
\eea
If there are $2N+1$ sites so that the chain ends with $A_{N+1}$ on the right boundary, then the BC becomes
\bea
&\;& \hskip -3.1cm B_{-1} = B_0 =0; \cr
&\;& \hskip -3.1cm  B_{N+1} = A_{N+2} =0.
\eea
Following similar steps carried out previously, we may again find the corresponding characteristic equation for $s$:
\bea  \label{Characteristic eq 1A odd}
&\;& \hskip -2.6cm  \left\{ U_{N+2}(u_1) U_{N}(u_2) - 2 U_{N+1}(u_1) U_{N+1}(u_2) + U_{N}(u_1) U_{N+2}(u_2) +2 \right\} \cr
&\;& \hskip -3.1cm  - \left(k_2/k_0 \right) \left\{ U_{N+1}(u_1) U_{N-1}(u_2) - 2 U_{N}(u_1) U_{N}(u_2) + U_{N-1}(u_1) U_{N+1}(u_2) +2 \right\} = 0.
\eea
Since the unit cell by the right boundary differs from that on the left, the corresponding winding numbers are different.  If fact, they are related by $\n_{\rm R} =1 - \n_{\rm L}.$

For a finite chain of type 2 extended SSH coupled-spring system with fixed-end BC, the Hamiltonian is given by
\bea
&\;& \hskip -0.0cm H_{\rm SSH-ext2}=\sum_{j=2}^{N-1}\Biggl\{ \frac{m}{2} \left[ \dot{a}_j^2(t) + \dot{b}_j^2(t) \right] + \frac{k_0}{2} \left[ a_j(t) - b_j (t) \right]^2 +  \frac{k_1}{2} \left[ a_{j+1}(t) - b_j(t)  \right]^2  +  \frac{k_{-1}}{2} \left[ a_{j-1}(t) - b_j(t)  \right]^2 \Biggr\} \cr
&\;& \hskip 2.7cm + \frac{m}{2} \left[  \dot{a}_{1}^2(t) + \dot{b}_{1}^2(t) \right] + \frac{k_0}{2} \left[ a_{1}(t) - b_{1} (t) \right]^2 + \frac{k_1}{2} \left[ a_{2}(t) - b_{1} (t) \right]^2 + \frac{k_{1} }{2} a_1^2(t) + \frac{k_{-1} }{2} b_1^2(t) \\
&\;& \hskip 2.7cm + \frac{m}{2} \left[  \dot{a}_{N}^2(t) + \dot{b}_{N}^2(t) \right] + \frac{k_0}{2} \left[ a_{N}(t) - b_{N} (t) \right]^2 + \frac{k_{-1}}{2} \left[ a_{N-1}(t) - b_{N} (t) \right]^2 + \frac{ k_{-1}}{2} a_N^2(t)  + \frac{ k_1}{2} b_N^2(t). \nn
\eea
The EOM:
\bea \label{SSH_ext2-R-semi-inf}
&\;& \hskip -3.1cm \Omh^2 A_j + \left(\kh_0 B_j + \kh_1 B_{j-1} + \kh_{-1} B_{j+1} \right) =0; \cr
&\;& \hskip -3.1cm \Omh^2 B_j + \left(\kh_0 A_j + \kh_1 A_{j+1} + \kh_{-1} A_{j-1} \right) =0,
\eea
where $\Omh^2 = \Om^2 - \left(\kh_0 + \kh_1 + \kh_{-1} \right).$
The BC becomes
\bea
&\;& \hskip -3.1cm A_0 =B_0 =0,  \quad A_{N+1}=B_{N+1}=0.
\eea
Similarly, we have the secular equation
\bea
&\;& \hskip -2.1cm \Omh^2 = \pm \sqrt{ \kh_0^2 + \kh_1^2 + \kh_{-1}^2 + 2\kh_1 \kh_{-1} \left(1 + 2 u_1 u_2 \right)  }.
\eea
The characteristic equation of $s$ is given by
\bea  \label{Simplified characteristic eq 1B even}
&\;& \hskip -2.6cm   \left\{ U_{N+2}(u_1) U_{N}(u_2) - 2 U_{N+1}(u_1) U_{N+1}(u_2) + U_{N}(u_1) U_{N+2}(u_2) +2 \right\} \cr
&\;& \hskip -3.1cm  - \left[ \frac{k_1}{k_{-1}} + \frac{k_{-1}}{k_1} \right]  \left\{ U_{N+1}(u_1) U_{N-1}(u_2) - 2 U_{N}(u_1) U_{N}(u_2) + U_{N-1}(u_1) U_{N+1}(u_2) +2 \right\}  \cr
&\;& \hskip -3.1cm +   \left\{ U_{N}(u_1) U_{N-2}(u_2) - 2 U_{N-1}(u_1) U_{N-1}(u_2) + U_{N-2}(u_1) U_{N}(u_2) +2 \right\}  = 0,
\eea
where $u_1$ and $u_2$ are related by
\bea
&\;& \hskip -3.1cm   u_1 +u_2 = - k_0\left(k_1 + k_{-1}\right)/(2 k_1 k_{-1} ). \nn
\eea
If there are $2N+1$ sites in the type 2 case, the system may be mapped to that of type 1 given in eq.~(\ref{Characteristic eq 1A odd}) under space inversion. So we will not repeat it here.

A direct check of the validity of the above results may be achieved by comparing the results obtained by diagonalizing the Hamiltonian with those obtained by solving the characteristic equation of $s$.  We will not show them here since similar results have been given in Ref. \cite{Ext-SSH}.  Suffice it to say they agree with each other to a very high accuracy.

\end{document}